\documentclass{llncs}
\usepackage{comment}
\usepackage{xcolor,graphicx,float}
\usepackage[normal]{subfigure}
\usepackage{dcolumn}
\usepackage{bm}

\usepackage{bbm}
\usepackage{algpseudocode}

\usepackage{algorithmicx,algorithm}

\usepackage{physics}
\usepackage{hyperref}
\usepackage[square,sort,comma,numbers]{natbib}

\usepackage{mathtools}
\DeclarePairedDelimiter\floor{\lfloor}{\rfloor}

\usepackage[capitalise]{cleveref}

\newtheorem{thm}{Theorem}
\crefname{thm}{Theorem}{Theorems}
\newtheorem{defn}{Definition}
\crefname{defn}{Definition}{Definitions}
\newtheorem{lem}{Lemma}
\crefname{lem}{Lemma}{Lemmas}
\newtheorem{conj}{Conjecture}
\crefname{conj}{Conjecture}{Conjectures}
\begin{document}

\title{Quantum approximate algorithm for NP optimization problems with constraints}

\author{Yue Ruan  \inst{1}\thanks{yue\_ruan@163.com} \and Samuel Marsh\inst{2} \and Xilin Xue \inst{1} 
\and Xi Li \inst{3}\and Zhihao Liu\inst{3} \and Jingbo Wang\inst{2}\thanks{jingbo.wang@uwa.edu.au}}

\institute{
School of Computer Science and Technology, Anhui University of Technology, Maanshan China
\and
School of Physics, University of Western Australia, Perth, Australia
\and
School of Computer Science and Engineering, Southeast University, Nanjing China
}

\maketitle              

\begin{abstract}
The Quantum Approximate Optimization Algorithm (QAOA) is an algorithmic framework for finding approximate solutions to combinatorial optimization problems, derived from an approximation to the Quantum Adiabatic Algorithm (QAA). In solving combinatorial optimization problems with constraints in the context of QAOA or QAA, one needs to find a way to encode problem constraints into the scheme. In this paper, we formalize different constraint types to linear equalities, linear inequalities, and arbitrary form. Based on this, we propose constraint-encoding schemes well-fitting into the QAOA framework for solving NP combinatorial optimization problems. The implemented algorithms demonstrate the effectiveness and efficiency of the proposed scheme by the testing results of varied instances of some well-known NP optimization problems. We argue that our work leads to a generalized framework for finding, in the context of QAOA, high-quality approximate solutions to combinatorial problems with various types of constraints.
\end{abstract}
\section{Introduction}

The Quantum Approximate Optimization Algorithm (QAOA), introduced by Farhi in 2014 \citep{farhi2014quantum}, is an algorithmic framework derived from an approximation to the Quantum Adiabatic Algorithm (QAA). The QAA \cite{farhi2001quantum}, published in 2001, has been intensively studied and is regarded as a promising quantum computing model. This is partly because of its ability to solve combinatorial optimization problems and its resistance to environmental noise and decoherence \cite{albash2018adiabatic}.

The QAA assumes a chosen quantum system can be adiabatically evolved from an easily-prepared initial Hamiltonian to a final ``problem'' Hamiltonian. The ground state of this final Hamiltonian encodes the solution to a combinatorial optimization problem. The adiabatic theorem guarantees that the system will remain in the instantaneous ground state provided the adiabatic condition is satisfied.  The system can then be measured after an appropriate length of time to obtain the optimal solution. 

QAOA approximates the adiabatic evolution by application of the Suzuki-Trotter theorem \cite{hall2015lie}. The time-dependent QAA Hamiltonian is broken into a sequence of $2p$ time-independent Hamiltonians. Each of these  time-independent operators have the form $\hat{U}_B(\beta)=e^{-i\beta \hat{B}}$ and $\hat{U}_C(\gamma)=e^{-i\gamma \hat{C}}$, where $\beta$ and $\gamma$ are tunable parameters, $\hat{C}$ is a diagonal problem Hamiltonian that encodes the solution qualities, and $\hat{B}$ can be considered as a ``mixing Hamiltonian''. The total evolution is produced by alternating applications of these two operators,
\begin{equation}\label{e1}
\ket{\vec\beta, \vec\gamma}=\hat{U}_B(\beta_p) \hat{U}_C(\gamma_p)...\hat{U}_B(\beta_1) \hat{U}_C(\gamma_1) \ket{s},
\end{equation}
where $\ket{s}$ is some initial state.
The optimal values for the set of parameters $\vec{\gamma}$ and $\vec{\beta}$ are those that maximize the expectation value $\bra{\vec\beta, \vec\gamma}\hat{C}\ket{\vec\beta, \vec\gamma}$. Intuitively, this is because a higher expectation value with respect to the problem Hamiltonian implies a higher average solution quality when the system is measured.

Compared to the QAA, the QAOA has greater flexibility in the design of the $\hat{B}$ operator (corresponding to the initial Hamiltonian in the QAA) and  $\hat{C}$ operator (corresponding to the final Hamiltonian in the QAA), due to the relaxed requirement on the energy gap between the ground and first excited state. 
However, both the QAOA and the QAA face the challenge of imposing problem constraints. A common method for encoding constraints is to add ``penalty'' terms to the problem Hamiltonian, whenever a constraint is violated \cite{lucas2014ising}. These terms add extra energy scales to the problem Hamiltonian, which in practice would increase the difficulty of getting the final solution \cite{hen2016quantum}. Since QAOA is an approximation scheme, there is a non-zero probability of producing a sub-optimal solution which may violate one or more constraints. It would be desirable to guarantee feasible solutions (which satisfy the problem constraints), and therefore the energy-penalty scheme is not appropriate.

In order to eliminate the nuisance of penalty terms, Hen present a scheme to encode the constraint in the initial Hamiltonian \cite{hen2016quantum, hen2016driver}. They find a constraint-encoded Hamiltonian $H_m$, which commutes with the initial Hamiltonian $\hat{B}$ but does not commute with the final Hamiltonian $\hat{C}$ -- that is, $[H_m, \hat{B}]=0$ but $[H_m, \hat{C}]\neq 0$. Exploiting this idea, they present effective operators to encode the constraints involved in problems such as graph partitioning, graph coloring, and not-all-equal 3-SAT. 

Hadfield inherited Hen's idea, extended their strategy to design sophisticated constraint-encoded $\hat{B}$ for a number of optimization problems in the context of QAOA \cite{hadfield2017qantum,hadfield2017quantum2,hadfield2018quantum}. Hadfield argue that encoding constraint in $\hat{B}$ rather than adding penalty terms in $\hat{C}$ will limit the size of the subspace under consideration so as to improve upon the average solution quality produced by QAOA. 

Marsh and Wang take a quantum walk assisted approach to encoding constraints in $\hat{B}$ under the framework of QAOA \cite{marsh2018quantum}. They consider the adjacency graph representation of the canonical transverse field Hamiltonian, and disconnect edges joining feasible solutions to unfeasible solutions. They limit their study to problems for which there is always a path from one feasible solution to all other feasible solutions on the transverse field Hamiltonian.

We study a variety of problems with constraints in the context of QAOA and classified the constraints into three categories -- linear equality constraint, linear inequality constraint, and constraints that do not fit into the former two categories. We find that given a single constraint that fits into a certain linear category, a property shared by many NP optimization problems, the constraint-encoding operator $\hat{B}$ can be represented by regular or near regular graphs, which are found to provide higher quality solutions than previous methods \cite{hadfield2017qantum,hadfield2017quantum2,hadfield2018quantum}.

Considering the adjacency graph representation of the operator $\hat{B}$, the graph should connect all feasible solutions and exclude unfeasible solutions. For general constraints, we propose a simple but effective ``star graph'' operator, which connects all feasible solutions with the center initial solution supporting any number of arbitrary constraints.

In the following sections, we first describe the prerequisite of our scheme, i.e. the combinatorial optimization problems should fall into the class of NP optimization (NPO) problems. Then we demonstrate our unified scheme for ``linear-form'' constraints, providing illustrative examples for well-known NP optimization problems as well as our ``star graph'' scheme to deal with arbitrary constraints. Finally we discuss the practicality of encoding constraints in the operator $\hat{C}$ rather than $\hat{B}$, such that the guarantee of measuring a feasible solution is preserved. 

\section{Prerequisites}
\label{Sec:Prerequisite}
In this paper we focus on NP optimization problems. By the definition of an NP optimization problem \cite{hromkovivc2013algorithmics}, we have access to an efficient oracle function $validate$ that can determine whether a given solution satisfies the constraints of the problem. Based on this, we define the following.
\begin{defn}
Let the set of feasible solutions be
\[ \Omega = \{ x \in 0 \ldots 2^n - 1 \mid validate(x) = 1 \}, \]
\end{defn}
where the integer number $x\in 0\cdots 2^n-1$ can be interpreted as an $n$ bits/qubits whose components `0/1' represents the solution to $n$ combinatorial optimization variables. And $\Omega$ can be obtained in the following way. Prepare a superposition state containing all combinatorial solutions, and add an auxiliary qubit with initial value $\ket{0}$. Then, by applying the quantum oracle that maps $\ket{x}\ket{0} \rightarrow \ket{x} \ket{validate(x)}$, we can select all feasible solutions in $\Omega$ with a controlled operation on the ancilla:
\begin{equation}\label{eq:omega}
\sum_{x=0}^{2^n-1}|x\rangle\otimes|0\rangle\xrightarrow{validate}\sum_{x \notin  \Omega}|x\rangle\otimes|0\rangle +\sum_{x \in \Omega}|x\rangle\otimes|1\rangle
\end{equation}

The solution $x$ is a $n$ bit string, where 1 indicates the corresponding combinatorial variable is included in the solution, while 0 refers to not. For the sake of eliminating the ambiguity, from now on we use $n$ dimensional vectors $\vec{x}$ to represent the solution. The $k-th$ component of $\vec{x}$ is represented as $x_k$. In addition, we assume the constraints of the NP optimization problem under consideration (inequality or equality) can be expressed as a polynomial function of $n$ binary variables $f(x_1,...,x_k,...,x_n)$. Since each $x_k \in \{0,1\}$, clearly $(x_k)^n=x_k$ and the interaction terms $x_1 * x_2 * \ldots * x_k$ have an effect only if $x_1=x_2=\dots=x_k=1$. One can use an extra variable $x_{\mu}$ to replace $x_1\cdots x_k$, rewrtie $\vec{c_{\mu}}*(x_1\cdots x_k)$ as $\vec{c_{\mu}}*x_{\mu}$. Then the polynomial constraint function $f$ can be simplified to
\begin{equation} \label{eq:f}
f(\vec{x})=\sum_{k} \vec{c}_k * x_k
\end{equation}
where $\vec{c}_k$ is the vector coefficient of $x_k$, defined below.
\begin{defn}
The coefficient $\vec{c}_k$ of $x_k$ is a vector of length $\kappa \in  \mathbbm{Z^+}$, that is, $\vec{c}_k=(c_{k1}, c_{k2},\dots, c_{k\kappa})$. The value of $\kappa$ differs depending on the problem under consideration. We define a linear equality constraint as one having the form $\sum_{k} \vec{c}_k * x_k=\vec{b}$, and a linear inequality constraint as $\vec{a} \leq \sum_k \vec{c}_k * x_k \leq \vec{b}$.
\end{defn}
Note that the constraint function $f$ takes $n$ binary variables (combinational variables) as entries, and output a $\kappa$ dimensional vector in general. In the extreme case, this vector could be degenerated to a scalar as in Graph Partition problem studied in Sect. 3.2.

\section{Scheme for linear equality constraints}\label{Sec-equality}
\subsection{Scheme 1}\label{sec:schemeequality}

At the beginning of this section, we would like to present a view of QAOA from the perspective of quantum walks. The Hamiltonian $\hat{B}$ can be observed as a graph; thus the unitary operator $e^{-i\hat{B}t}$ can be regarded as a continuous quantum walk on it \cite{childs2004quantum}. The walker can reach all the connected nodes and the walking is restricted in the range defined by the connected subgraph to which the starting node belongs. Based on this understanding, encoding constraint in $\hat{B}$ is equivalent to find a way  connecting all feasible solutions (nodes) and excluding all infeasible solutions (nodes). Hence, we have the following theorem.

\begin{thm}\label{thm:t1}
For the class of optimization problems whose constraint subjects to $\sum_k \vec{c}_k * x_k=\vec{b}$. If (1) all feasible solutions contain exactly the same number of ``1s'' (2) one feasible solution is trivially known, and from it, we can find a ``0-1'' swapping sequence to traverse all feasible solutions, then the constraint encoded $\hat{B}$ can be constructed as:
\begin{equation}\label{eq:equalityb}
\hat{B}=\sum\limits_{\substack{\vec{x}, \vec{x'} \in \Omega \\ d(\vec{x},\vec{x'})=2}}\ket{x}\bra{x'}+\ket{x'}\bra{x},
\end{equation}
where $d(\vec{x},\vec{x'})$ is the Hamming distance between solutions $\vec{x}$ and $\vec{x'}$.
\begin{proof}
As we know, the solution to the optimization problem is a bit string. According to conditions (1) and (2),  all the solutions contain the same number of  ``1s'', but they differ in where the ``1s'' appear. From the trivial solution selected, all other feasible solutions can be generated by successive ``0-1'' swapping operations. Therefore, the permutation can be created by varied swapping positions and swapping orders. That is to say, if the trivial solution selected is $\vec{x_0}$, an arbitrary feasible solution $\vec{x_{t}}$ can be obtained by a permutation $\tau$, that generates a feasible solution sequence $\{\vec{x_0}, \vec{x_1}, \cdots, \vec{x_t}\}$, meeting $\tau(\vec{x_0})=\vec{x_1}, \tau(\vec{x_1})=\vec{x_2}, \cdots, \tau(\vec{x_{t-1}})=\vec{x_{t}}$.

Select feasible solutions $\vec{x}$ and $\vec{x'}$ randomly in this sequence, meeting $\tau(\vec{x}) = \vec{x'}$. We know that $\vec{x}$ and $\vec{x'}$ only differ in $2$ positions by one swapping between ``0'' and ``1'', so the Hamming distance between them is $2$. Therefore, $\vec{x}$ and $\vec{x'}$ are two neighbour feasible nodes on graph $\hat{B}$, their relationship can be described as $\vec{x}, \vec{x'} \in \Omega \land d(\vec{x}, \vec{x'})=2$. Exploiting operator $\ket{x}\bra{x'}+\ket{x'}\bra{x}$ to connect these two nodes (i.e. $B_{x,x'} \leftarrow 1$), then the entire constraint-encoded operator $\hat{B}$ can be gotten as per \ref{eq:equalityb}.
\end{proof}
\end{thm}

\begin{thm}
The operator given in \cref{eq:equalityb} can be efficiently constructed and the corresponding $\hat{U}_B(\beta)$ has an efficient quantum circuit.
\begin{proof}
We use results from \cite{aharonov2003adiabatic}, which gives a method for implementation of $\hat{U}_B(\beta)$ providing that $\hat{B}$ is sparse and efficiently row-computable. This is indeed the case for $\hat{B}$ as per \cref{eq:equalityb}. Consider the following pseudo-code for generating $\hat{B}$:
\begin{algorithmic}[1]
\State $\vec{x} \gets$ row to be computed \qquad /* i.e. $\sum\limits_{\vec{x} \in \Omega}\ket{x}$ */
\For{$\vec{x'} \in \{\vec{x'}|d(\vec{x'}, \vec{x}) = 2\} $}
    	\State $\vec{x'} \gets$ flip $2$ bits of $\vec{x}$
    	  \If {$validate(\vec{x'})$} 
        \State $B_{x,x'} \leftarrow 1$
    \EndIf
\EndFor
\end{algorithmic}

The number of flipping in the \textbf{for} loop should be bounded by $\mathcal{O}(\tbinom{n}{2})$, i.e. $\mathcal{O}(n^2)$. And since the $validate$ function is by definition efficient, each row of $\hat{B}$ can be efficiently generated. Given that $\vec{x} \in \Omega$ (each row) is in a superposition state, the above operation can be done simultaneously to construct the entire $\hat{B}$ efficiently. 

This pseudo-code also implies that $\hat{B}$ is sparse, since the number of non-zero elements per row is bounded by $n^{2}$. Hence, exploiting Aharonov's decomposition lemma, $\hat{B}$ can be decomposed as $poly(n)$ $2\times2$ combinatorially block diagonal matrices so that an efficient quantum circuit can simulate $\hat{U}_B(\beta)$ within the desired accuracy.\cite{aharonov2003adiabatic}.
\end{proof}
\end{thm}

Note that Theorem 1 solves optimization problems with constraint stated as $\sum_k \vec{c}_k * x_k=\vec{b}$ as well as additional conditions (1) and (2). These conditions imply that the feasible solution space has a certain degree of symmetry. Many NP optimization problems, including partition problems, packing problems, and scheduling problems, exhibit such a structure of feasible solutions. In the following sections, we will study the effectiveness of the constraint-encoding operator given in \cref{eq:equalityb} through these problems.

\subsection{Example 1 --- Graph Partition}
\label{Sec:GP}
The first problem is graph partitioning as studied in \cite{lucas2014ising,hen2016quantum}. It can be described as: Given a graph with even vertices, the graph partition problem is to find a half-and-half vertices partition such that the number of edges connecting two subsets is minimized. Let $n$ be a multiple of two so the vertices can be split evenly into two sets, and let $\vec{x} = x_n \ldots x_1$ represent the solution where $x_i = 1$ if vertex $i$ is in the first set and $0$ if in the second. The operator $\hat{C}$ for this problem is defined as $\hat{C}=|E|-\frac 1 2 \sum_{(u,v)\in E}(1-\sigma_u^z\sigma_v^z)$ (E is the edge set of the given graph), such that a solution $\vec{x}$ has a higher solution quality when there are fewer edges connecting the two subsets \cite{difference}. The constraint is that each subset should have the same number of vertices, so $\sum_k x_k=n/2$. This is the simplest form required by \cref{thm:t1}, in which the coefficient $\vec{c_k}$ of each $x_k$ has degenerated to a scalar $c_k$ with a value of  ``1'' and $\vec{b}$ to a scalar $n/2$. For this problem, we divide the first half vertices in subset $1$ and the remaining in subset $0$, which denotes a trivial feasible solution ``$11\cdots100\cdots0$". It is easy to verify as long as we continue swapping the ``0'' and ``1''  once each time, we can obtain all the other possible solutions starting from this trivial solution. Therefore, we can specify operator $\hat{B}$ by the following equation.
\begin{equation}\label{e5}
\hat{B}=\sum\limits_{\substack{\vec{x}, \vec{x'} \in \Omega \\ d(\vec{x},\vec{x'})=2}}\ket{x}\bra{x'}+\ket{x'}\bra{x}
\end{equation}

\begin{figure}[htbp]\label{Fig1}
    \subfigure[An example graph partition problem instance.]{
    \label{Fig1-instance}
    \includegraphics[width=11cm]{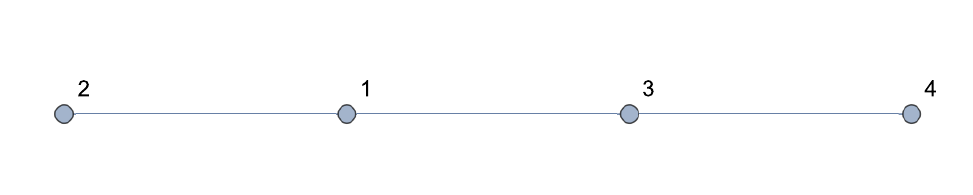}
    }  
    \quad  
    \subfigure[The structure of our operator as per \cref{e5}.]{
    \label{Fig1-OurMethod}
    \includegraphics[width=11cm]{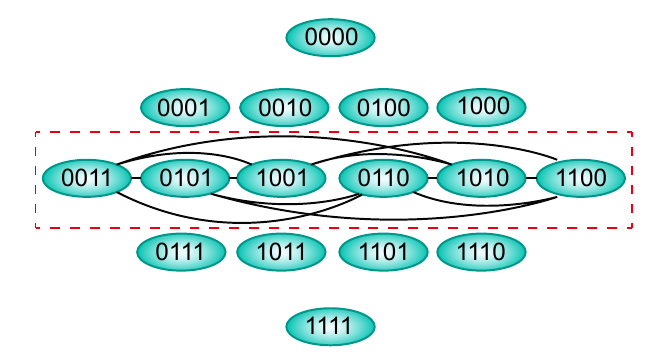}   
    }
    \quad  
    \subfigure[The structure of our operator as per \cref{e6}.]
	{ 
    \label{Fig1-TheirMethod}
    \includegraphics[width=11cm]{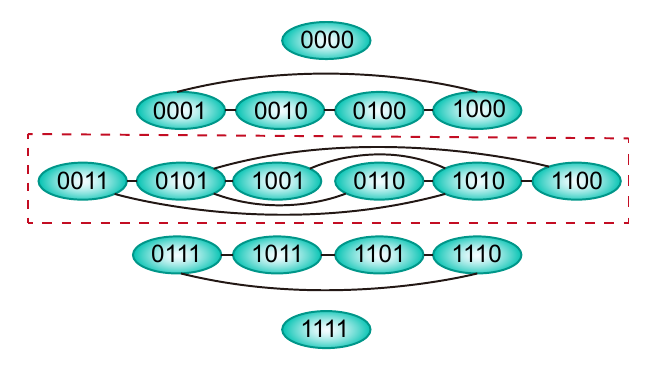}  
    }  
    \caption{The structure of different constraint-encoding operators $\hat{B}$, for the graph partition problem.}
\end{figure}

\cref{Fig1-instance} is the specific instance we choose for graph partition. Our corresponding constraint-encoding operator $\hat{B}$ has structure illustrated in \cref{Fig1-OurMethod}. We can see that the nodes in the middle level, framed by the red dotted line, construct a connected subgraph corresponding to constraint-satisfied subspace.

For the problem of graph partition, Hadfield and Hen defined $\hat{B}$ as
\begin{equation}\label{e6}
\hat{B}=-\sum_{i=1}^n(\sigma_i^x \sigma_{i+1}^x+\sigma_i^y \sigma_{i+1}^y)
\end{equation}
which encodes the constraint. Applying this to the graph shown in \cref{Fig1-instance}, it will generate multiple isolated connected subgraphs based on the number of  ``1s'', as shown in \cref{Fig1-TheirMethod}. The subgraph in the middle level corresponds to the constraint-satisfied subspace of feasible solutions. For this problem, only the middle subspace is consequential.

\begin{figure}[htbp]  
    \subfigure[The probability distribution obtained using our $\hat{B}$ as per  \cref{e5}.]{ 
	\label{Fig: probability1}
    \includegraphics[width=11cm]{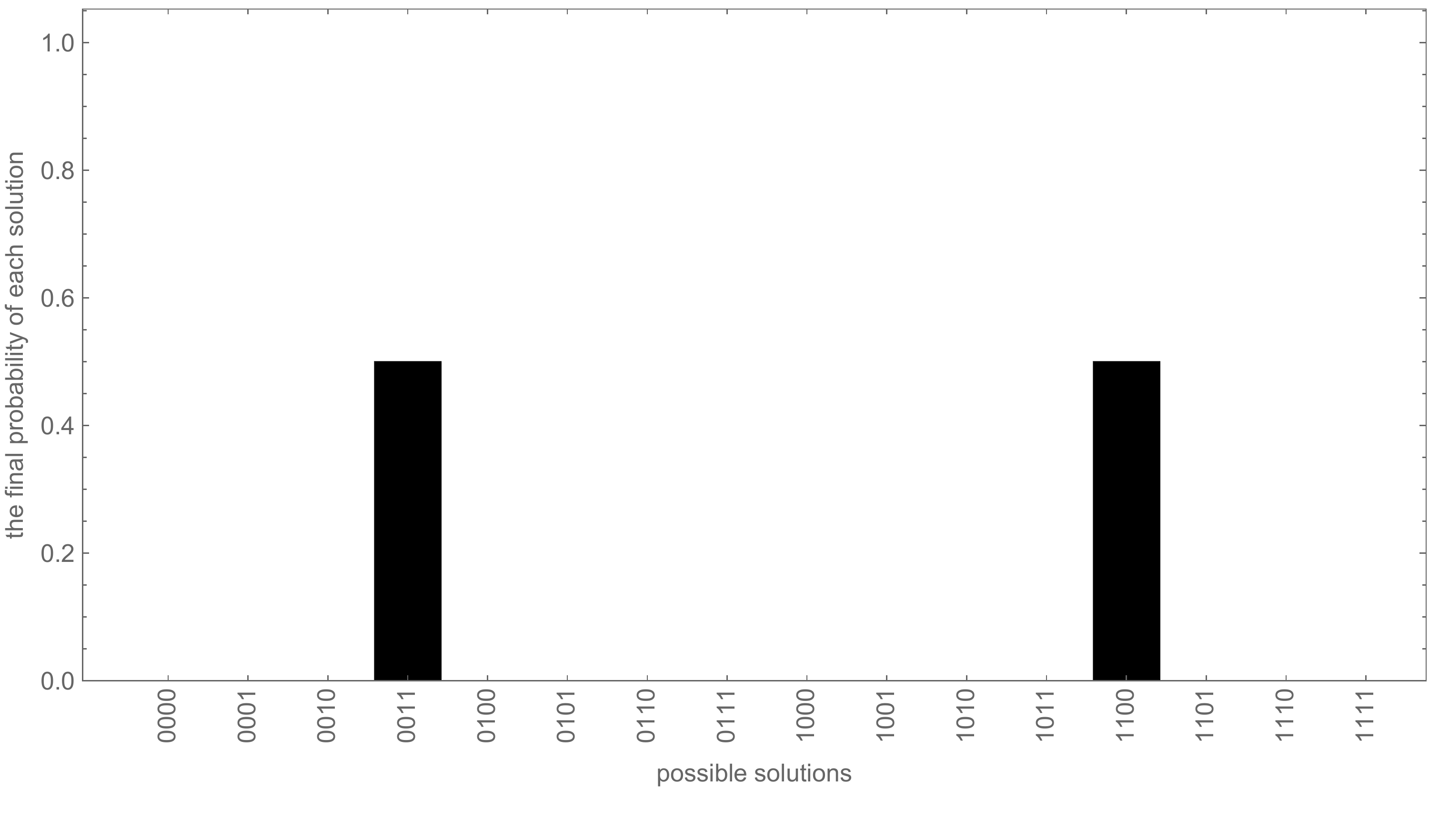}  
     }  
    \quad  
    \subfigure[The probability distribution obtained using Hen's $\hat{B}$ as per \cref{e6}.]{  
	\label{Fig: probability2}
    \includegraphics[width=10cm]{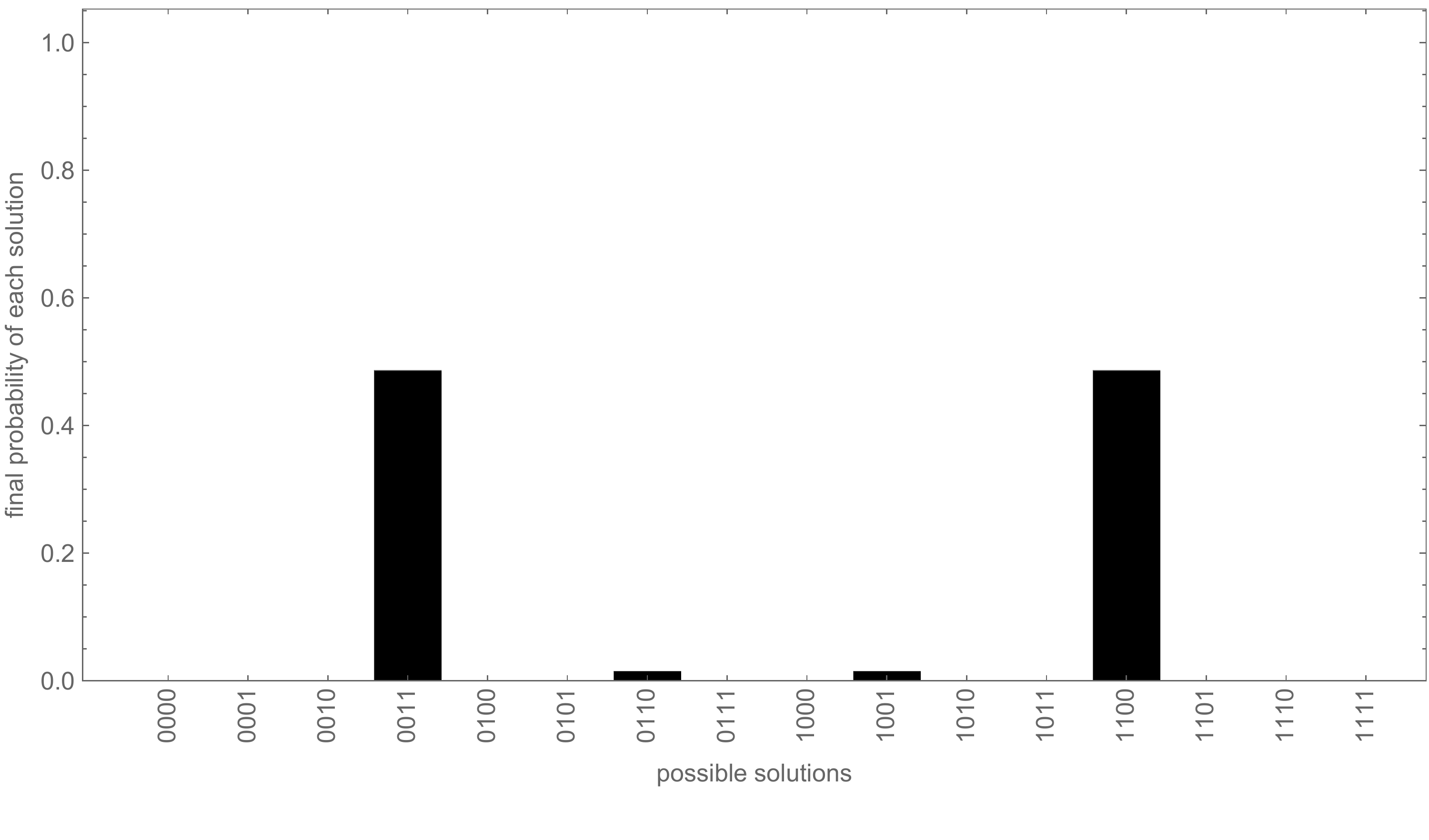}  
    }  
    \caption{The results of graph partition for the problem of graph partitioning on instance \cref{Fig1-instance}, using QAOA with $p=3$. The bit-string in the horizontal axis represents a solution to the problem (possible partition). The value in the corresponding vertical axis represents the probability of getting such solution.}
    \label{fig:GraphPartitioningResults}
\end{figure}  

One difference between \cref{Fig1-OurMethod} and \cref{Fig1-TheirMethod} is that our scheme does not generate redundant subgraphs. Moreover, the two feasible-solution subgraphs have different connectivity structures. The \cref{Fig1-OurMethod} subgraph is 4-regular. In contrast, the \cref{Fig1-TheirMethod} subgraph has 2 nodes with degree 4 and 4 nodes with degree 2. Such a difference will lead to a different probability distribution at the end of QAOA. In \cref{fig:GraphPartitioningResults}, the bit-string in the horizontal axis represents a solution to the problem (possible partition). For example, ``0011'' means vertices ``1'' and ``2'' in subset $1$, while vertices ``3'' and ``4'' in subset $0$. The value in the corresponding vertical axis represents the probability of getting such solution. Obviously, our method has a perfect result (\cref{Fig: probability1}), while the other method introduces a small amount of noise, thus producing a small probability of obtaining a sub-optimal solution (\cref{Fig: probability2}). The comparison suggests that regularity in the $\hat{B}$ -- that is, not biasing any feasible solution vertex over any other -- is a positive factor for obtaining an improved probability distribution. In \cref{sec:symmetry} we provide evidence for this conjecture using larger instances of graph partition and other combinatorial problems.

\subsection{Example 2 --- Multiple processor scheduling}
\label{Sec-MPS}
The Multiple Processors Scheduling problem can be described as follows. We are given $n$ tasks, each with processing times $t_k$ for $k = 1 \ldots n$, and we have $m$ processors. The problem at hand is to find an assignment of tasks to processors, so that the time needed to accomplish these tasks is minimized. 

We use $mn$ qubits to represent the solution space, with solutions defined using $x_i = 1$ if and only if the $(i \text{ mod } n)$'th task is run on the $\floor{i/n}$'th processor. For example, $\ket{11\ldots1,00\ldots0,\ldots,00\ldots0}$ is a trivial feasible solution where all tasks run on the first processor.

The problem Hamiltonian can be defined as
\begin {equation}
\hat{C}=-\sum\limits_{x=0}^{2^{mn} - 1} \max\Big\{\sum_{k=i n}^{(i + 1)n - 1}t_{(k\ mod\ n)} x_k, i = 0 \ldots (m - 1) \Big\} \ket{x} \bra{x}
\end{equation}

Naturally, the constraint is that each task should be scheduled exactly once, and only on a single processor. A feasible solution $\vec{x} = x_0x_1\dots x_{mn-1}$ for $m$ processors should meet
\begin{equation}
\sum\limits_{i=0}^{m-1} x_{i n + j} = 1
\end{equation}
for all tasks $j = 0 \ldots (n - 1)$. We can express this in the form
\begin{equation}
\sum_{k=0}^{mn-1} \vec{c}_k * x_k = (1, 1, \ldots, 1)^T
\end{equation}
where each coefficient $\vec{c}_k$ of $x_k$ is a length-$n$ vector with elements
\begin{equation}
(\vec{c}_k)_j = \begin{cases}
1 & k \text{ mod } n = j \\
0 & \text{otherwise}
\end{cases}
\end{equation}
for $j = 0 \ldots (n - 1)$. From the above trivial example $\ket{11\ldots1,00\ldots0,\ldots,00\ldots0}$ selected, other feasible solutions can be traversed by successive ``0-1'' swapping at the corresponding location (processor), namely ``$x_i \leftrightarrow x_{i +n} \leftrightarrow \cdots \leftrightarrow x_{i +(m-1)n}$'' $(i =0,\cdots, n-1)$.  This exactly fits the form required by \cref{thm:t1}. Hence, \cref{e5} is an appropriate choice for the $\hat{B}$ operator.

Here, we give a concrete example for two processors and five tasks $\{A, B, C, D, E\}$ with running times $\{3, 4, 8, 2, 5\}$. The result is shown in \cref{mutli-processor}. The two solutions with highest probability correspond to tasks A and C scheduled on one processor, with the rest scheduled on the other. It is straightforward to verify that this is the optimal solution.

\section{Scheme for linear inequality constraints}\label{Sec-inequality}

\subsection{Scheme 2}\label{sec:schemeinequality}

Our method for handling linear inequality constraint is similar to that in \cref{sec:schemeequality}. We simply change the Hamming distance of connected solutions from $2$ to $1$. That is,
\begin{equation}\label{eq:bsingleinequality}
\hat{B}=\sum\limits_{\substack{\vec{x}, \vec{x'} \in \Omega \\ d(\vec{x}, \vec{x'})=1}}\ket{x}\bra{x'}+\ket{x'}\bra{x}.
\end{equation}
As with the previous operator, $\hat{U}_B(\beta)$ also has an efficient quantum circuit for $\hat{B}$ as per \cref{eq:bsingleinequality}, and the proof is much the same as discussed above.

The following theorem will describe the scope of this scheme. We again assume constraint function $f$ can be written in the linear form of \cref{eq:f}.
\begin{thm}\label{t2}
Let $f$ be subject to $\vec{a}\leq \sum_{k} \vec{c}_k * x_k \leq \vec{b}$, where $(\vec{c}_{k})_\mu \geq 0$ and $b_\mu - a_\mu \geq 2\max_k\{(\vec{c}_{k})_\mu\}$ for all components $\mu$. Then all feasible solutions are connected using the operator given in \cref{eq:bsingleinequality}.

\begin{lem} The feasible solutions form a non-strict partial order $\Omega$.
\begin{proof}
Let non-strict partial order $\preceq$ on the set of bit-strings be $\vec{x} \preceq \vec{x'}$ if and only if every bit in $\vec{x}$ is less than the corresponding bit in $\vec{x'}$. We can verify that $\vec{x}$ has the following properties: (for convenience, define matrix $C=(\vec{c_1}\cdots \vec{c_n})$, $C \cdot \vec x= \vec{c_1}*x_1+ \cdots +\vec{c_n}*x_n$):

\begin{enumerate}
\item Reflexivity: $\vec x \preceq \vec x \Longleftrightarrow C \cdot \vec x \leq C \cdot \vec x$,
\item Anti-symmetry: $\vec{x'} \preceq \vec{x}$ and $\vec{x} \preceq \vec{x'}$ $ \Longleftrightarrow C \cdot \vec{x'} \leq C \cdot \vec{x}$\ and $C \cdot \vec{x} \leq C \cdot \vec{x'}$  $\Longleftrightarrow C \cdot \vec{x} = C \cdot \vec{x'}$ $\Longleftrightarrow \vec x = \vec {x'}$,
\item Transitivity: $\vec {x''} \preceq \vec {x'}$ and $\vec {x'} \preceq \vec {x}$ $\Rightarrow C \cdot \vec {x''} \leq C \cdot \vec {x'}$\ and $C \cdot \vec {x'} \leq C \cdot \vec {x}$  $\Rightarrow C \cdot \vec {x''} \leq C \cdot \vec {x}$ $\Rightarrow \vec {x''} \preceq \vec {x}$.
\end{enumerate}
  
Hence, the set $\Omega$ containing all feasible $\vec{x}$ is a non-strict partial order set.
\end{proof}
\end{lem}
\begin{lem} For all $\vec{x}, \vec{x'} \in \Omega$, there exists a path connecting $\vec{x}$ and $\vec{x'}$ on the operator defined in \cref{eq:bsingleinequality}.
\begin{proof}
Regard $\sum\limits_{\substack{\vec{x}, \vec{x'}\in \Omega \\ d(\vec{x},\vec{x'})=1}}\ket{x}\bra{x'}+\ket{x'}\bra{x}$ as an adjacency matrix. The connectivity structure is the same as the Hasse diagram of partial order set $\Omega$ (substituting directed edges for undirected edges). 

Consider first the case where $\vec{x}\preceq \vec{x'}$ or $\vec{x'}\preceq \vec{x}$. The relationship between $\vec{x}$ and $\vec{x'}$ is indeed ancestor and descendant. So there must be a path which connects them on the Hasse diagram.
Consider now the case where $\vec{x}$ and $\vec{x'}$ are incomparable. We know that $\vec{a}\leq f(x_1,...,x_n)\leq \vec{b}$ and $b_\mu - a_\mu \geq 2\max_k\{(\vec{c}_{k})_\mu\}$. This condition guarantees that if we do one bit-flip $1\rightarrow 0$ at any place, or one bit-flip $0 \rightarrow 1$, at least one of these will produce a feasible solution as well. Therefore in the corresponding Hasse diagram, there are at least two complete levels. Two levels of the Hasse diagram are a fully connected subgraph. On this subgraph, there must be nodes which are the ancestor or descendant of $\vec{x}$ and $\vec{x'}$. So there must be edges, starting from these nodes, to connect $\vec{x}$ and $\vec{x'}$.
\end{proof}
\end{lem}
\begin{proof}
By combining Lemmas 1 and 2, we obtain the theorem as required.
\end{proof}
\end{thm} 

For the purpose of demonstrating the effectiveness and efficiency of operator $\hat{B}$ defined by \cref{eq:bsingleinequality}, we would like to use another famous NP-complete problem, Set Packing listed by Karp\cite{karp1972reducibility}, to show the detail.

\subsection{Example 1 --- Set Packing}

The Set Packing problem can be described as follows. There is a set $S$ and some subsets $S_1,S_2,...,S_n\subseteq S$. Can we find $k$ subsets which do not have elements in common? The corresponding optimization question is ``what is the maximum value of this $k$?'' The constraint can be formulated as: 
\begin{equation}
\bigcup\limits_{selected\  i, j}S_i\cap S_j=\emptyset \, .
\end{equation}

We now give a concrete instance. Consider $S=\{a_1, a_2, a_3, a_4, a_5, a_6\}$ and $S_1=\{a_1, a_3\}$, $S_2=\{a_2\}$, $S_3=\{a_4, a_5\}$, $S_4=\{a_2, a_5, a_6\}$. Clearly, the trivial solution meeting the constraint is to select no sets.  The optimal solution is to choose ($S_1, S_2, S_3$), which do not have elements in common. We can use a 4-bit string $\vec{x}=x_3x_2x_1x_0$ to represent the solution, with 0/1 at every position to represent the non-selection/selection of the corresponding subset respectively. Then we define the coefficients $\vec{c_0}=(000101)^T$, $\vec{c_1}=(000010)^T$, $\vec{c_2}=(011000)^T$ and $\vec{c_3}=(110010)^T$ to indicate the elements contained in each $x_k$ (subset). Therefore the constraint can be rewritten in the form defined by \cref{t2}:
 
\begin{equation}
\sum\nolimits_{k}\vec{c_k}*x_k=\vec{c_0}*x_0+\vec{c_1}*x_1+\vec{c_2}*x_2+\vec{c_3}*x_3\leq (111111)^T
\end{equation}

\begin{figure}[htbp]\label{fig:5}
    \subfigure[The corresponding graph of transverse field operator $\sum\limits_l \sigma_l^x$]{  
	\label{fig5-transverse}
    \includegraphics[width=11cm]{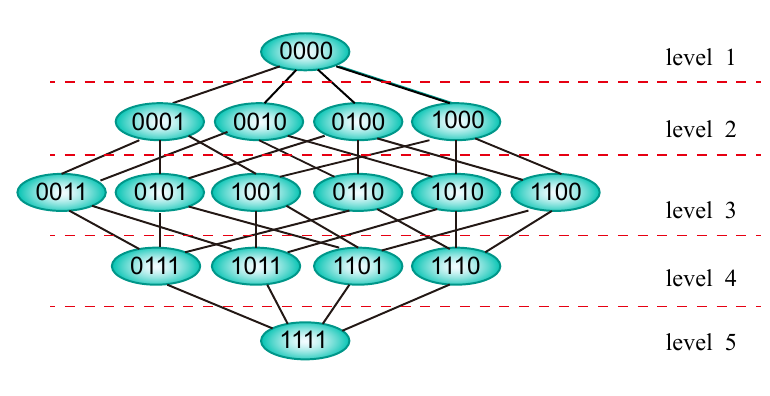}  
     }  
    \quad  
    \subfigure[The graph of tailored $\hat{B}$ for Set Packing (our method)]{  
	\label{fig5-our-tailored}
	\includegraphics[width=11cm]{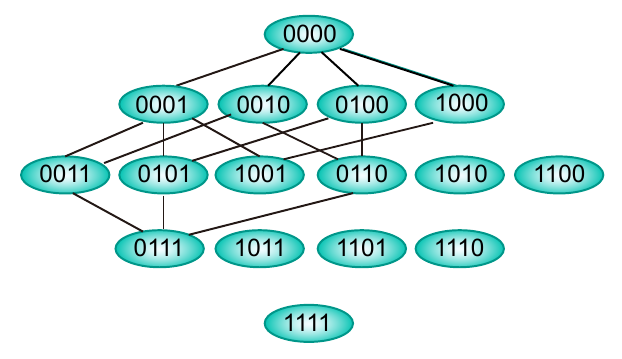}
    }  
     \caption{The graphical view of canonical $\hat{B}$ and tailored $\hat{B}$ }
\end{figure}

For a better illustration of the solving procedure, we present a graphical view of the canonical operator $\hat{B}$, i.e. transverse field Hamiltonian $H_0=\sum_l \sigma_l^x$. As shown in \cref{fig5-transverse}, the nodes correspond to the eigenvectors of $H_0$. For this Set Packing problem, they represent the selection/non-selection of 4 subsets. The nodes can be divided into several hierarchical levels based on the number of  ``1s'' in the bit-string. The edge connects nodes with pairwise Hamming distance 1, and generates a connected undirected graph.

We use a similar algorithmic procedure in \cref{sec:schemeequality} to efficiently row-compute our $\hat{B}$ operator, as follows:

\begin{algorithmic}[1]
\State $\vec{x} \gets$ row to be computed \qquad /* i.e. $\sum\limits_{\vec{x} \in \Omega}\ket{x}$ */
\For{$\vec{x'} \in \{\vec{x'}|d(\vec{x'}, \vec{x}) = 1\} $}
    	\State $\vec{x'} \gets$ flip $1$ bit of $\vec{x}$
	 \If {$validate(\vec{x'})$} 
        \State $B_{x,x'} \gets 1$
      \EndIf
\EndFor
\end{algorithmic}

The inner block is executed $n$ times, and since the $validate$ function is  by definition efficient, this $\hat{B}$ can be generated efficiently.
As can be seen in \cref{fig5-our-tailored}, every feasible node is connected and every unfeasible node is isolated. The feasible nodes construct a connected subgraph and we can find a path from the trivial feasible node ``0000'' to the optimal node ``0111''. 

Hadfield defines ``operator controlled-bit-flip mixers''
\begin{equation}
H_{CX,j}=2^{-D_j}\sigma_j^x\prod_{i\in nbhd(j)}(I+\sigma_i^z)
\end{equation}
to encode the constraint for Set Packing. In this equation, $i$ is the controlled-bit to swap subset $S_j$ in or out of the undetermined optimal set $S_0$ ($S_0$ is a set composed of disjoint subsets). The quality is determined by the mixing rule when trying to incorporate a new subset. So the complexity of the mixer operator will depend on the number of subsets, i.e. $\mathcal O(n)$. 

Although the form of two operators are different, the connectivity structure of the constraint-encoded subgraph is the same as shown in \cref{fig5-our-tailored}, so that they both can get an identical high-quality result, as shown in Fig 4. But we note that although the final result is the same, the idea behind each method is different.
\begin{figure}[!htbp]\label{Fig7}
    \includegraphics[width=11cm]{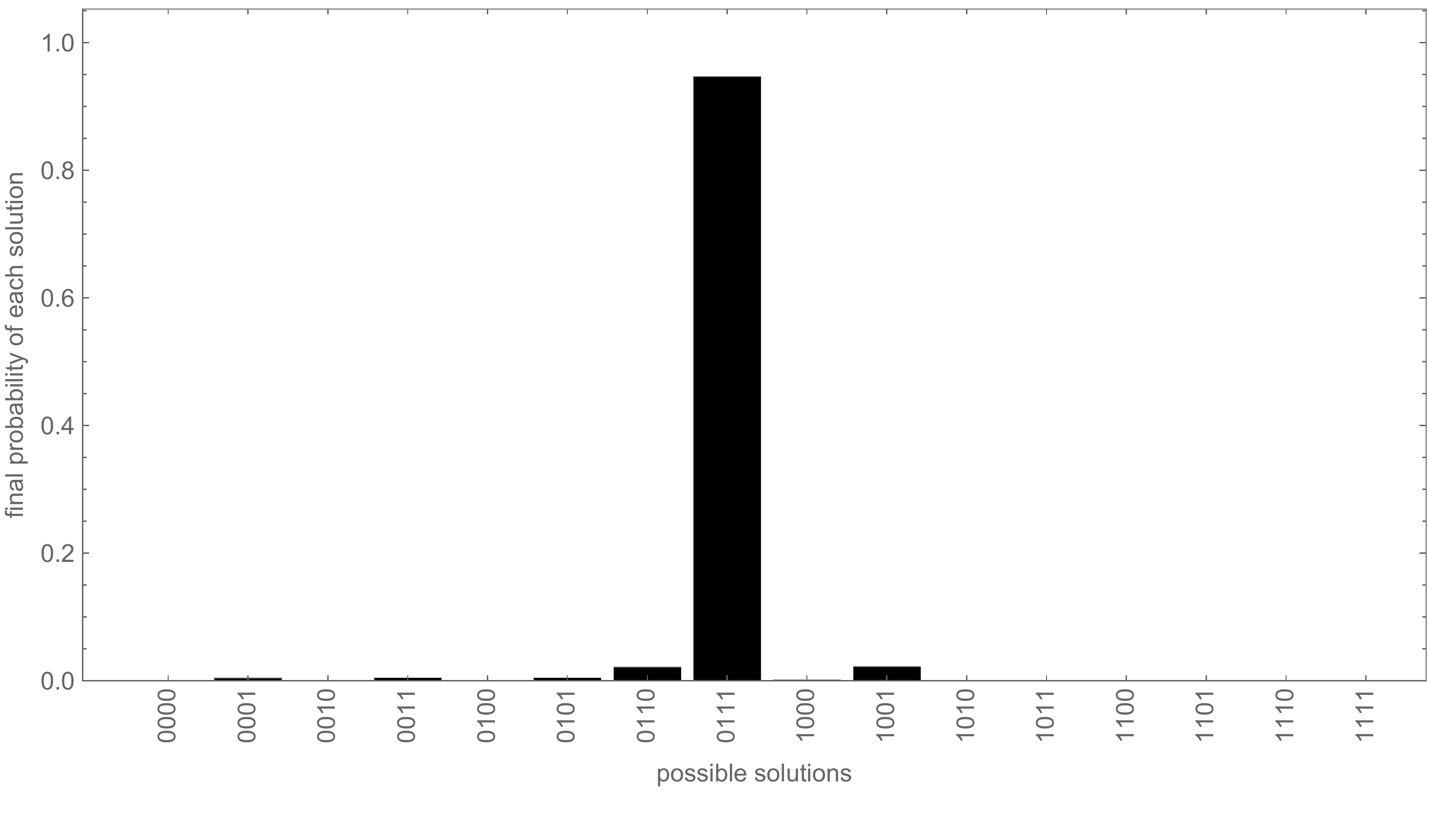}  
    \caption{The probability distribution for the Set Packing problem produced by QAOA.}
\end{figure}  

\subsection{Example 2 --- vertex cover}
We use another well-known vertex cover  problem to show the application of Theorem 3. A vertex cover is a set of vertices of a graph such that each edge is incident to at least one vertex of the set. The goal of the optimization problem of minimum vertex cover is to minimize the set size, given the constraint.

The problem Hamiltonian (operator $\hat{C}$) can be defined as:
\begin{equation}
\hat{C}=\sum_{x=0}^{2^n-1} \vec{x}\cdot\vec{e}\ket{x}\bra{x}
\end{equation}
where $\vec{x}=x_0x_1\dots x_{n-1}$, each $x_i=1$ or $0$ represents the selection or non-selection of $n$ vertices, and $\vec{e}=(11\dots1)^T$.

\begin{figure}[!htbp]
    \subfigure[The testing graph.]{ 
	\label{cover1}
    \includegraphics[width=9cm]{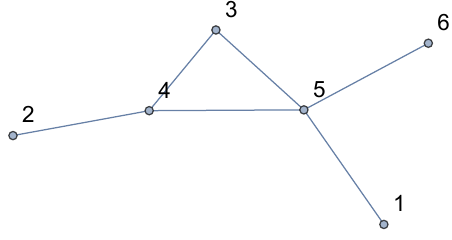}  
     }  
    \quad  
    \subfigure[The final probability distribution for the testing graph.]{  
	\label{cover2}
    \includegraphics[width=11cm]{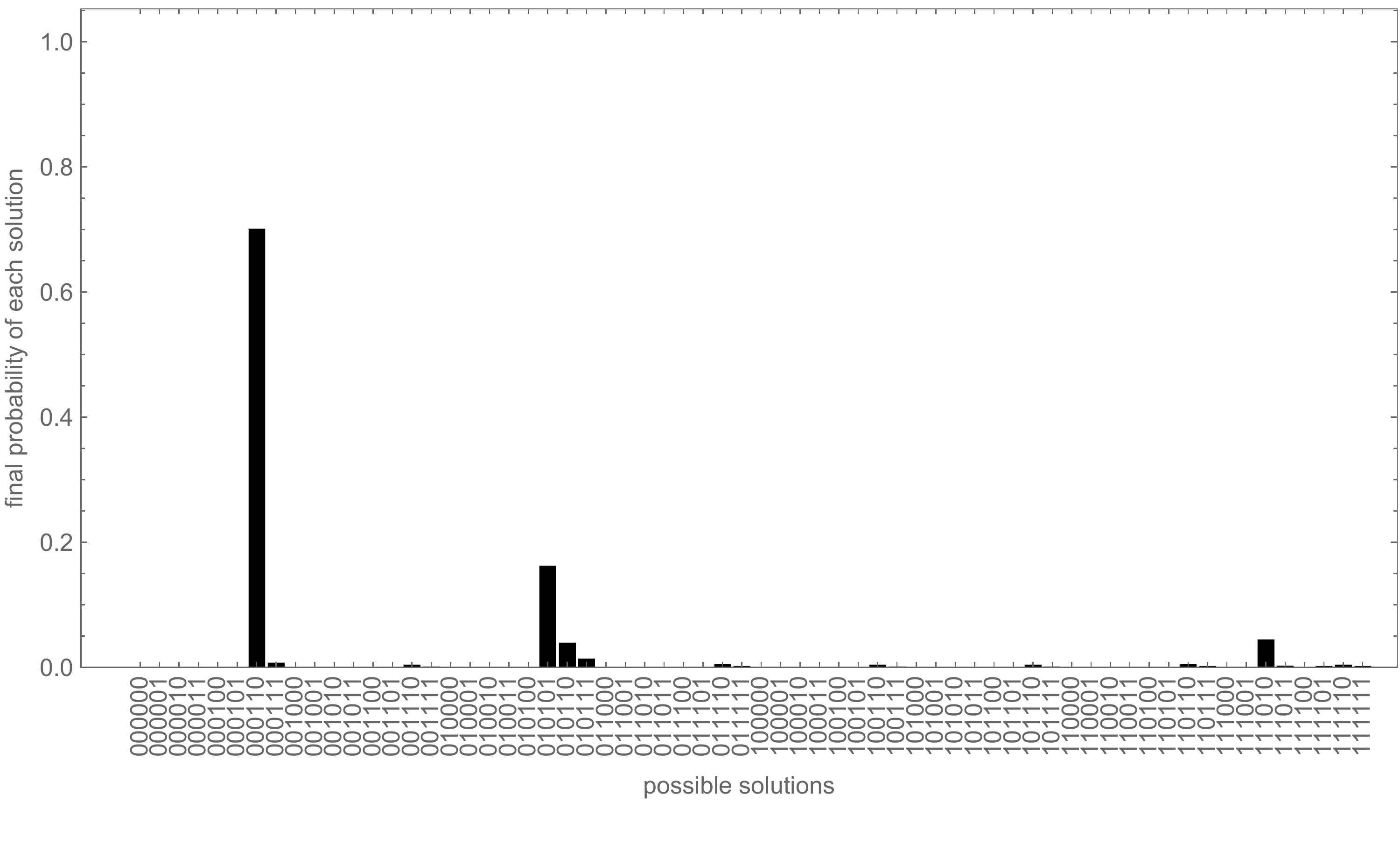}
    }  
    \caption{The minimum vertex cover of the testing sample under $p=3$.}
\end{figure}

The graph of $n$ vertices has possible edges $n(n-1)/2$. We can use an $n(n-1)/2$ bit string to map these edges. For example, the graph shown as \cref{cover1} can be mapped to (1-2, 1-3, 1-4, 1-5, \dots, 5-6), where 1-2 means a possible edge between the vertices 1 and 2. If the edge exists, that value is 1, and otherwise is 0. The coefficient $\vec{c_0}$ of $x_0$ (the possible edges of vertex 1) can be defined as a vector $(00010\dots0)^T$, which means vertex 1 is only associated with edge 1-5. Therefore, the constraint for \cref{cover1} can be constructed as the exact form of \cref{t2} as given by 
\begin{equation}\label{f11}
\sum\nolimits_{k}\vec{c_k}*x_k \geq (000100100110101)^T
\end{equation}
The result of running QAOA is shown in \cref{cover2}.

\section{Scheme for arbitrary constraints}
\subsection{Example --- Multiple processor scheduling with additional constraints}
\label{sec:mps}

As we recall, in \ref{Sec-MPS} Multiple Processor Scheduling was introduced as an example to show the constraint-encoded $\hat{B}$ (\ref{eq:equalityb}) to handle linear equality constraint. But in practice, there may be additional constraints on the scheduling. As an example, two tasks may have conflicts and may result in a bottleneck if they are running on the same processor. Alternatively, there may be some ordering dependence between two or more tasks where it is inappropriate to schedule them on different processors. In such scenarios, the constraint function $f$ may not have the linear form, so it is inappropriate to make use of our aforementioned schemes.

To show this clearly, we give another concrete instance: 3 tasks (A, B and C) schedule on 2 processors but we demand B and C should not run on the same processor. In that case just 4 nodes meet the constraints: ``110001'', ``010101'', ``101010'', ``001110''. It can be easily verified that all these nodes can neither be connected by a single rule $d(\vec{x},\vec{x'})=1\ or\ 2$ nor be handled by the single ``swap'' or ``partial swap'' strategy presented by Hadfield\cite{hadfield2017quantum2}.

So, how can we handle such a scenario?

\subsection{Star graph method}

For combinational problems falling into NP class, it is easy to find a trivial solution $\vec{x^*}$ meeting the constraint generally. Knowing this, we present a simple but effective ``star graph'' method to encode the constraint in operator $\hat{B}$. First, we choose an arbitrary feasible solution $\vec{x^*}$. Then we define the operator as

\begin{equation}
\label{eq:star}
\hat{B} = \sum_{\vec{x},\vec{x^*}\in \Omega} \ket{x}\bra{x^*}+\ket{x^*}\bra{x}
\end{equation}

We have to note one row in $\hat{B}$ is not row sparse in general (the row to which $\vec{x^*}$ belongs). Therefore, the approach in \cite{aharonov2003adiabatic} is not applied to simulating $e^{-i\beta\hat{B}}$. Fortunately, we know that $\hat{B}$ is a low rank Hamiltonian with $rank(\hat{B})=2$. In addition, $Tr(\hat{B})=1$ implies that $\hat{B}$ is a legal density operator. These two properties of operator $\hat{B}$ meet the condition of exploiting the efficient Hamiltonian simulation approach presented by Lloyd\cite{lloyd2014quantum}. The detailed simulation procedure can be found in \cite{kimmel2017hamiltonian}.

\begin{figure}[!htbp]
    \subfigure[\cref{eq:equalityb} approach.]{ 
	\label{mutli-processor}
    \includegraphics[width=11cm]{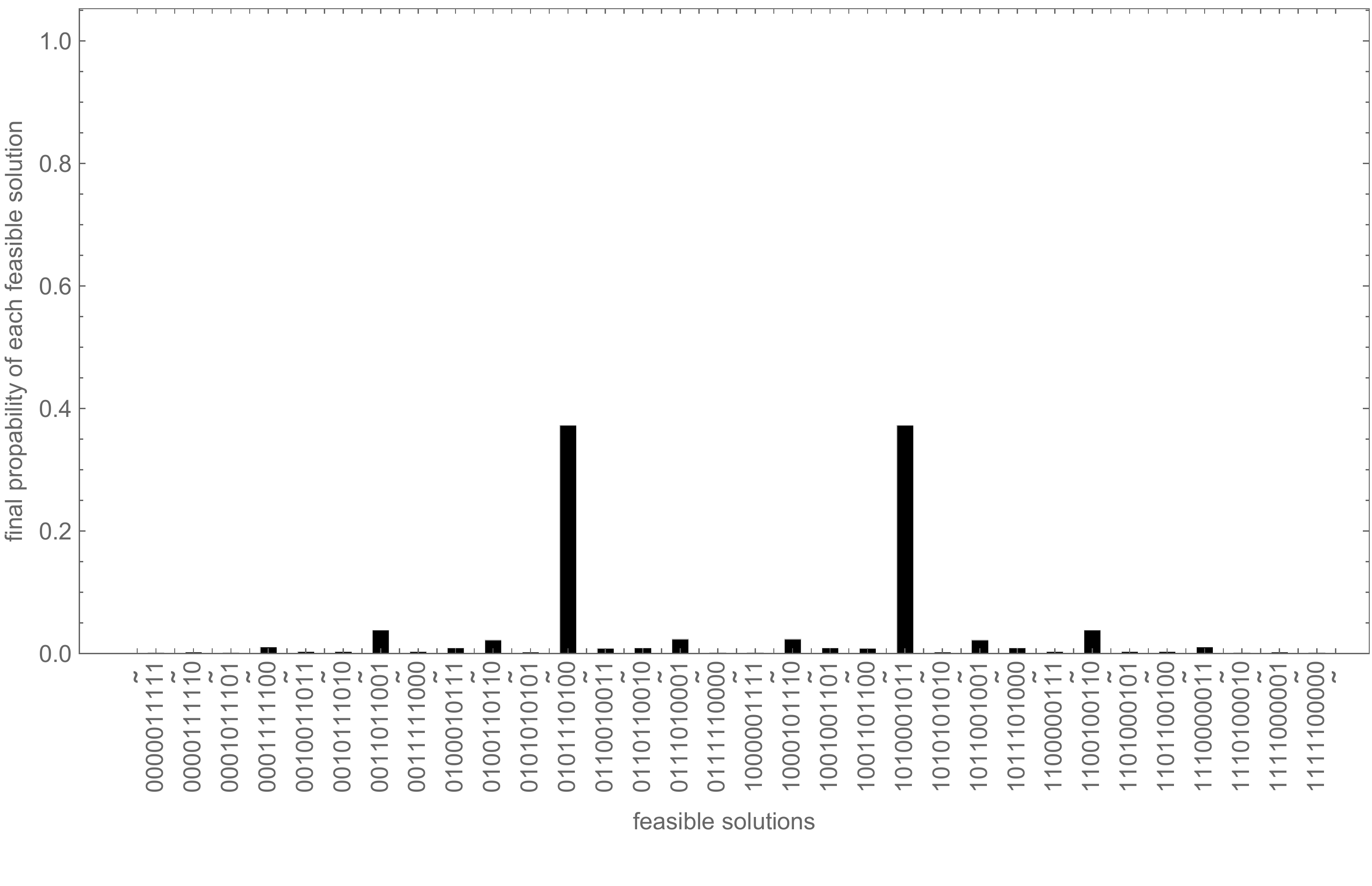}  
     }  
    \quad  
    \subfigure[Star graph approach, with extra constraints.]{  
	\label{Richardmutli}
    \includegraphics[width=11cm]{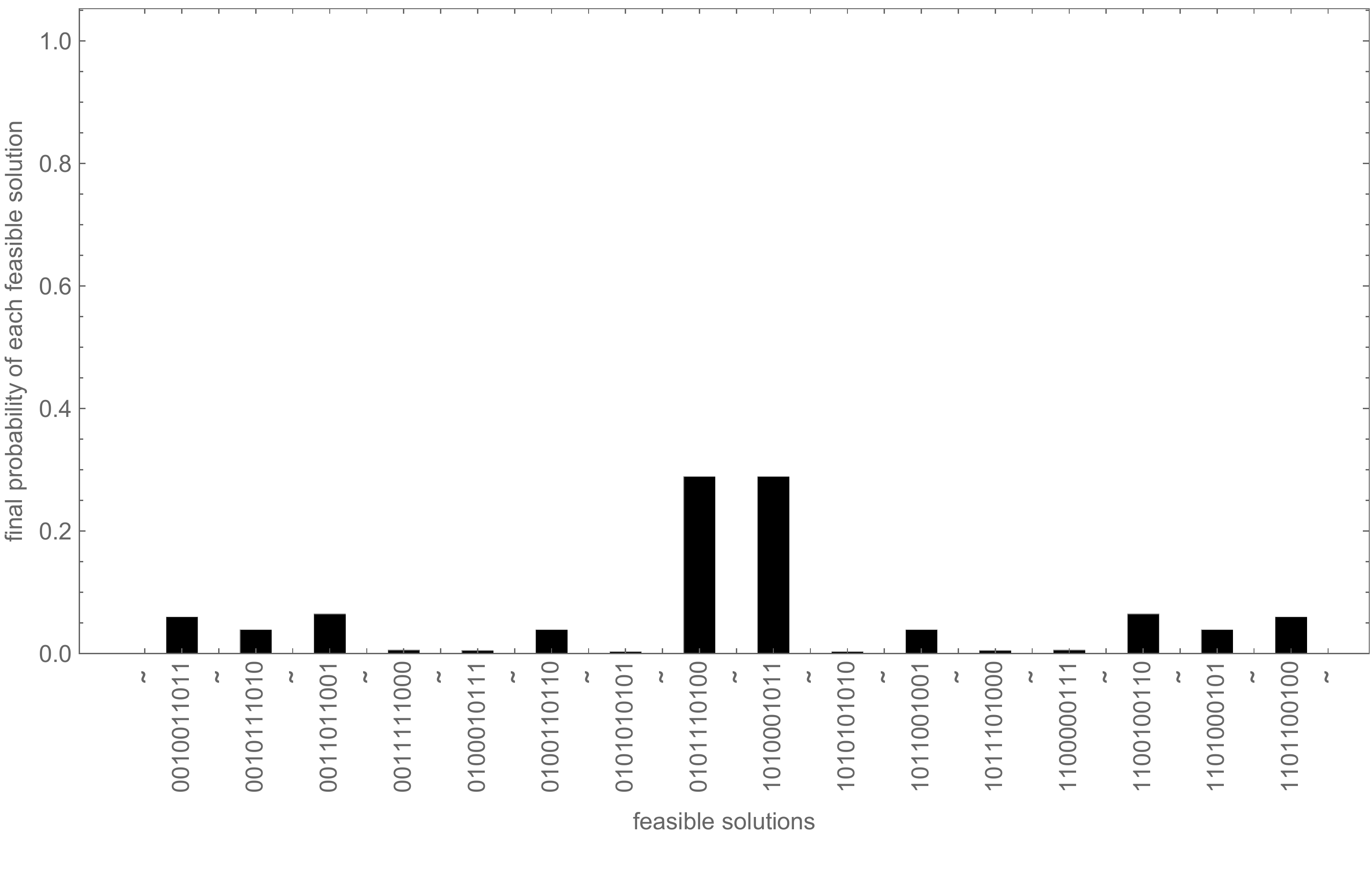}
    }  
    \caption{Results for Multiple Processor Scheduling.}
\end{figure}

We now apply this star graph method to the concrete problem instance given in Sect. \ref{Sec-MPS}, with the extra constraint that C and D should not be scheduled on the same processor. \cref{Richardmutli} shows the result. We can see that the solutions with the highest probabilities are ``0101110100'' and ``1010001011''. This represents A, B and D running on one processor, with C and E running on the other processor. Comparing this to \cref{mutli-processor}, we can see different optimal solutions emerging when additional constraints are added.

The star graph method is an effective method to encode \textit{arbitrary} constraints in operator $\hat{B}$, but we do not recommend it as the first choice. Using the star graph method, the central node $\vec{x^*}$ is assigned too much weight. That unbalanced structure leads to inferior performance. That is to say, the aforementioned methods should be prioritized if applicable to the problem at hand. The reason for such a choice is based on the conjecture that the regularity of operator $\hat{B}$ is a positive factor for achieving better performance.

\section{The effect of the symmetry in operator $\hat{B}$} \label{sec:symmetry}

\begin{conj}
In the context of using QAOA to solve combinational optimization problems with constraints, if the corresponding graph of the constraint encoded $\hat{B}$ has more regularity, then it is in favour of achieving a higher probability of the optimal solution than other constraint encoded $\hat{B}$ with less regularity under the condition of fixed p iteration.
\end{conj}

\subsection{Verification for equality constraints}

We use graph partition problem as the testing problem. We auto-generate a large number of random graphs with 6 vertices, and then compare the results of running QAOA with different $\hat{B}$. Specifically, we consider \cref{e5}, \cref{e6} (Hen's and Hadfield's method) and \cref{eq:star} (the star graph method with central node chosen randomly).

\begin{figure}[!htbp]
    \subfigure[The problem instance]{ \label{Fig:9a}
    \includegraphics[width=9cm]{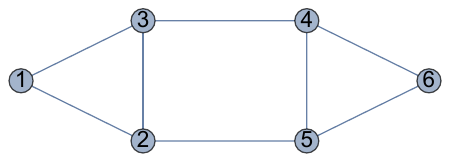}  
     }  
    \quad  
    \subfigure[\cref{e5} approach]{
	\label{Fig:9b} 
    \includegraphics[width=12cm]{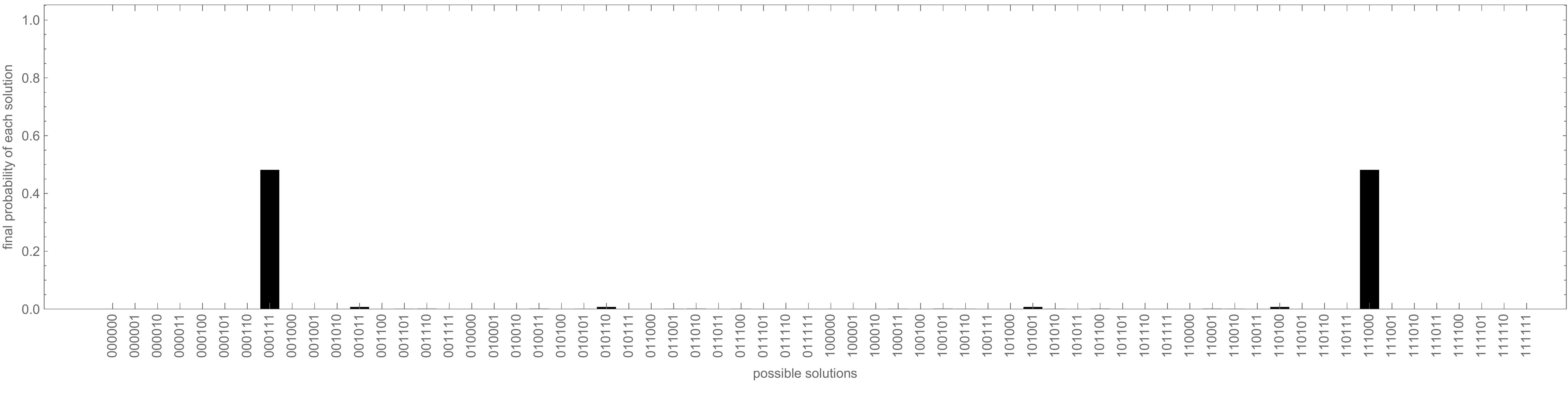}
    }  
    \quad  
    \subfigure[\cref{e6} approach]{  
	\label{Fig:9c} 
    \includegraphics[width=12cm]{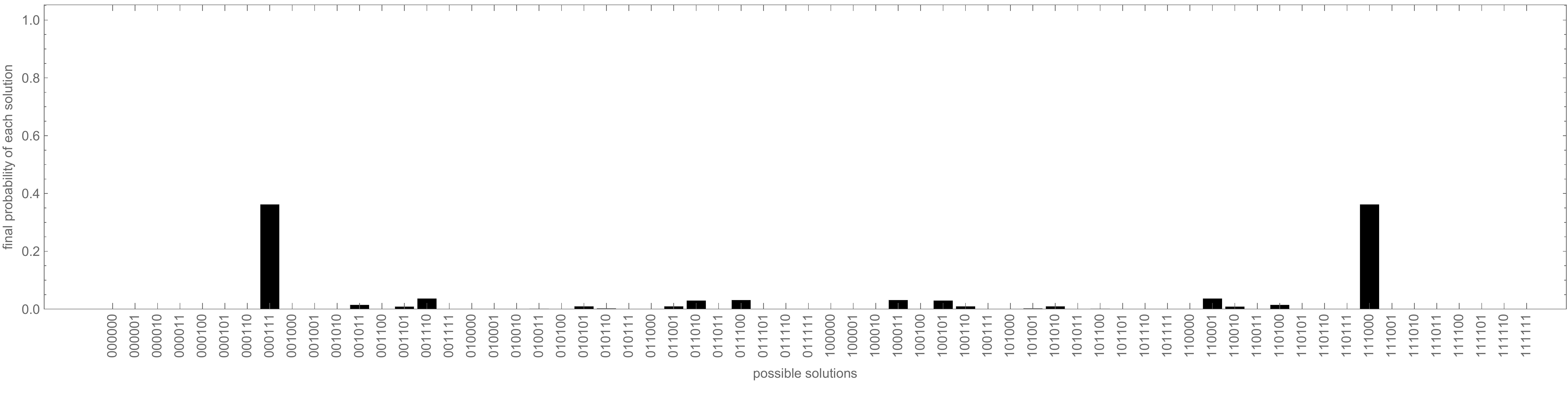}
    }  
    \quad  
    \subfigure[\cref{eq:star} approach, $\vec{x^*}=(010101)^T$]{  
	\label{Fig:9d} 
    \includegraphics[width=12cm]{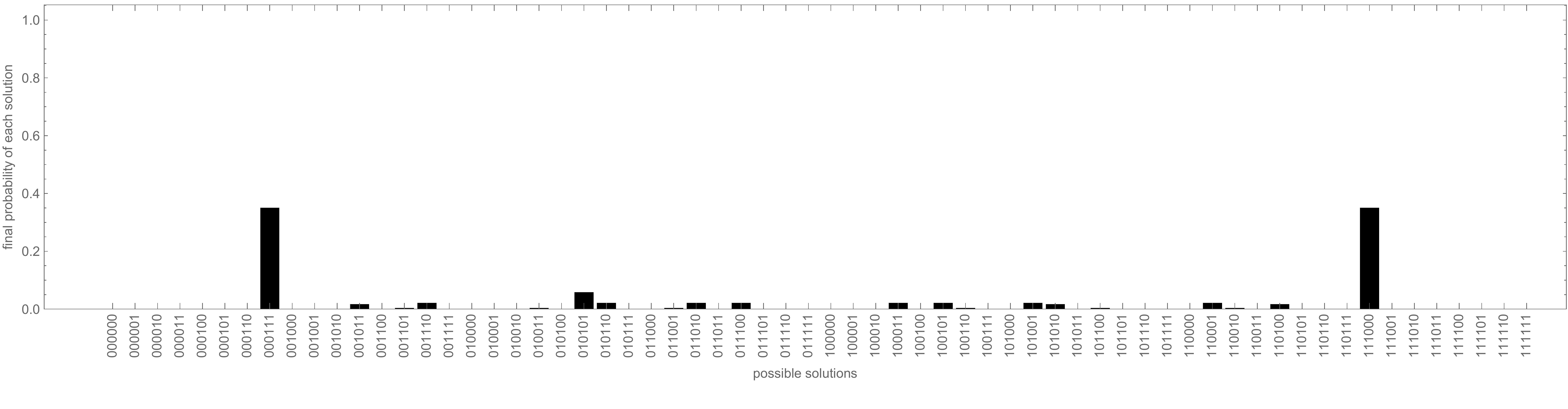}
    }
    \quad  
    \subfigure[\cref{eq:star} approach, $\vec{x^*}=(000111)^T$]{  
	\label{Fig:9e} 
    \includegraphics[width=12cm]{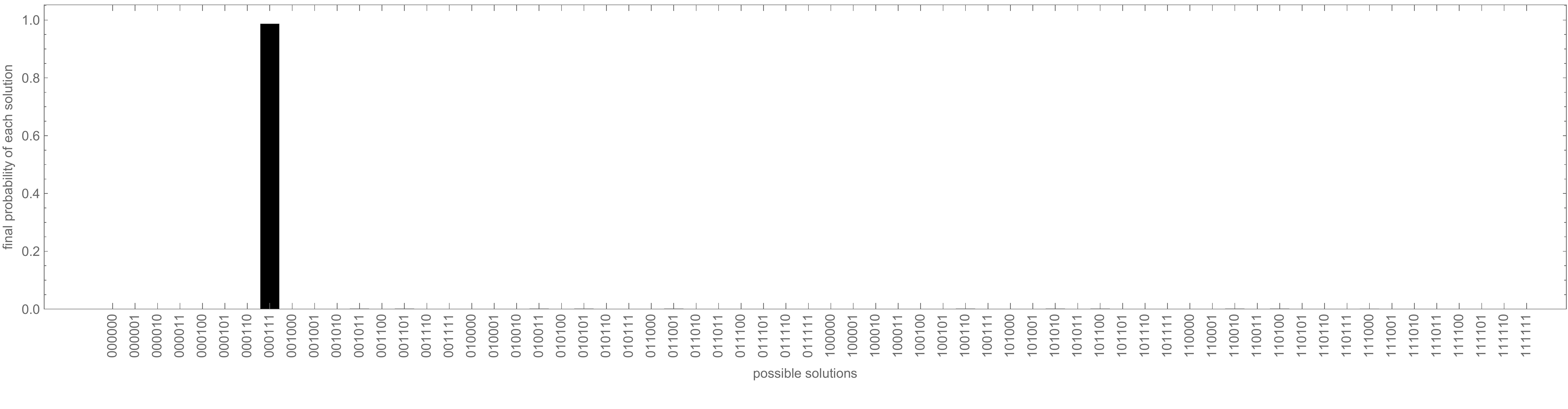}  
    }
    \caption{Results for the graph partition problem with $p=5$.}
\end{figure}

\begin{figure}[!htbp]
    \subfigure[The problem instance]{  \label{Fig:10a}
    \includegraphics[width=9cm]{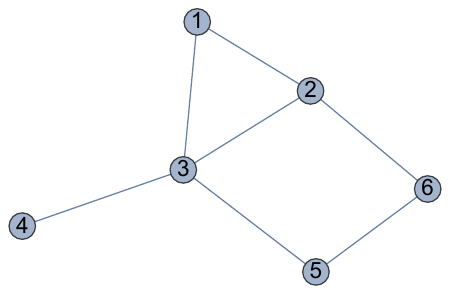}  
     }  
    \quad  
    \subfigure[\cref{e5} approach]{
	 \label{Fig:10b}  
    \includegraphics[width=12cm]{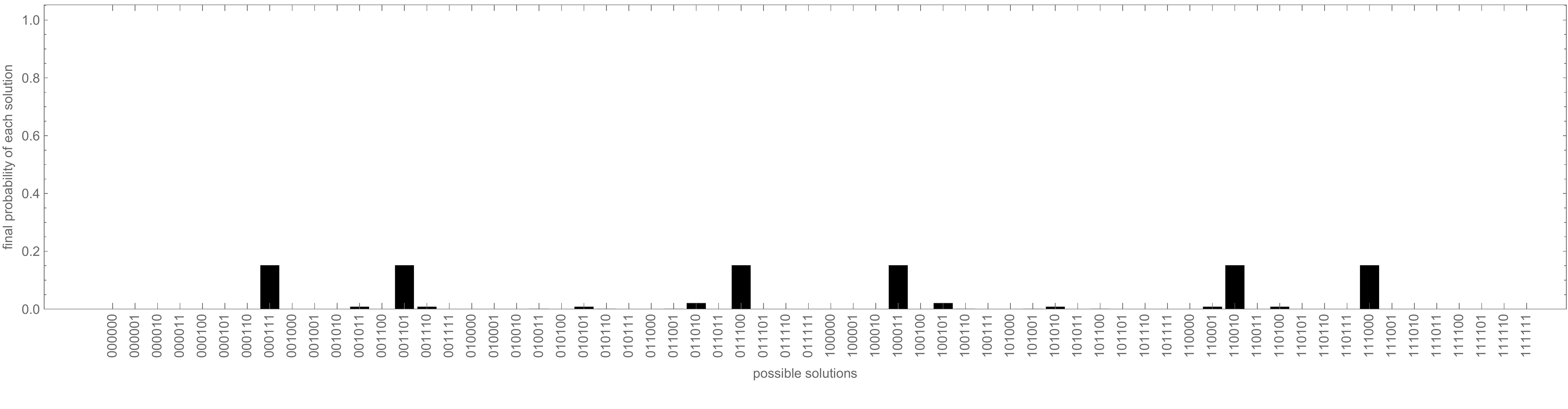}
    }  
    \quad  
    \subfigure[\cref{e6} approach]{
	 \label{Fig:10c}  
    \includegraphics[width=12cm]{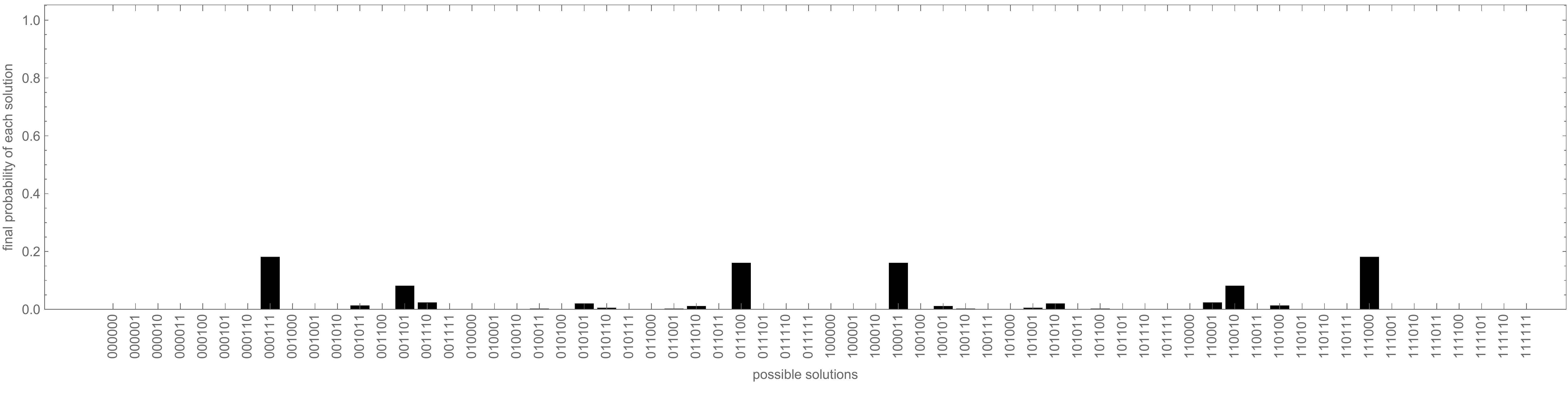}
    }  
    \quad  
    \subfigure[\cref{eq:star} approach, $\vec{x^*}=(010101)^T$]{  
	 \label{Fig:10d}
    \includegraphics[width=12cm]{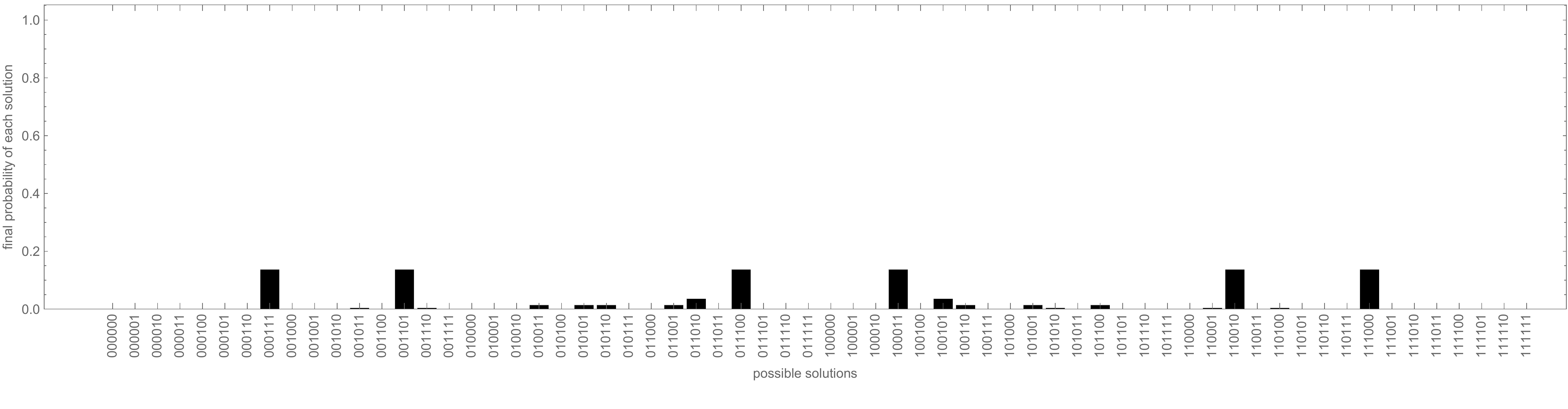}
    }
    \quad  
    \subfigure[\cref{eq:star} approach, $\vec{x^*}=(000111)^T$]{  
	 \label{Fig:10e}
    \includegraphics[width=12cm]{graphpartition6-21star.pdf}  
    }
    \caption{Results for the graph partition problem with $p=5$.}
\end{figure}

Without exception, $\hat{B}$ as defined by Eq. \ref{e5} are regular graphs. Indeed, the result of exploiting this operator is better than other methods. As representative examples, we select \cref{Fig:9a}, which has a single optimal partition, and \cref{Fig:10a}, which has three optimal solutions. From these figures, we can see that for the first instance \cref{e5} (\cref{Fig:9b}) produces the highest-quality result by a large margin. For the graph with more than one optimal partition, \cref{e5} (\cref{Fig:10b}) wins by a slim margin (the probability of one optimal partition in \cref{Fig:10c} is too low). The star graph operator has similar performance. However, we emphasize that the solutions of the star graph approach depend on the selected $\vec{x^*}$. If we happen to choose $\vec{x^*}$ as an optimal solution, other equally optimal solutions will be missed, as \cref{Fig:9e} and \cref{Fig:10e} show.

\subsection{Verification for inequality constraints}

\begin{figure}[!htbp] 
\label{Fig-9}
    \subfigure[\cref{eq:bsingleinequality} approach, $p=3$]{
	\label{Fig-9a}  
    \includegraphics[width=12cm]{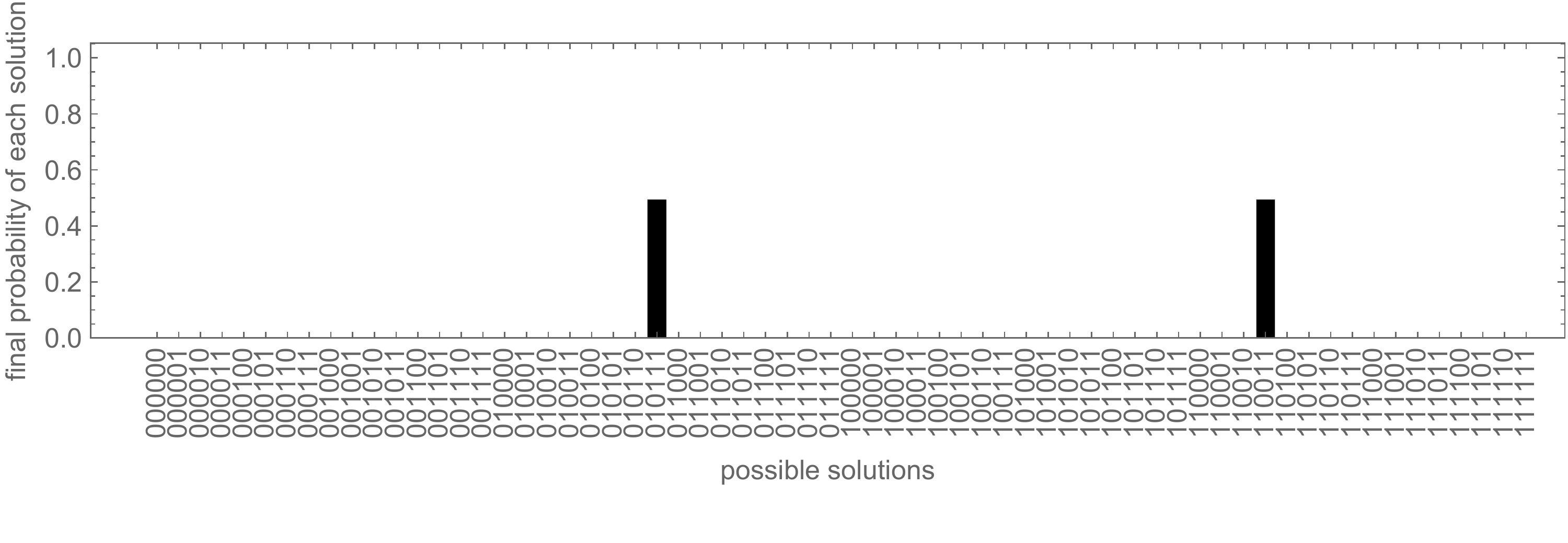}
    }  
    \quad  
    \subfigure[Star graph approach, $\vec{x^*}=(000000)^T$, $p=3$]{ 
	\label{Fig-9b}  
    \includegraphics[width=12cm]{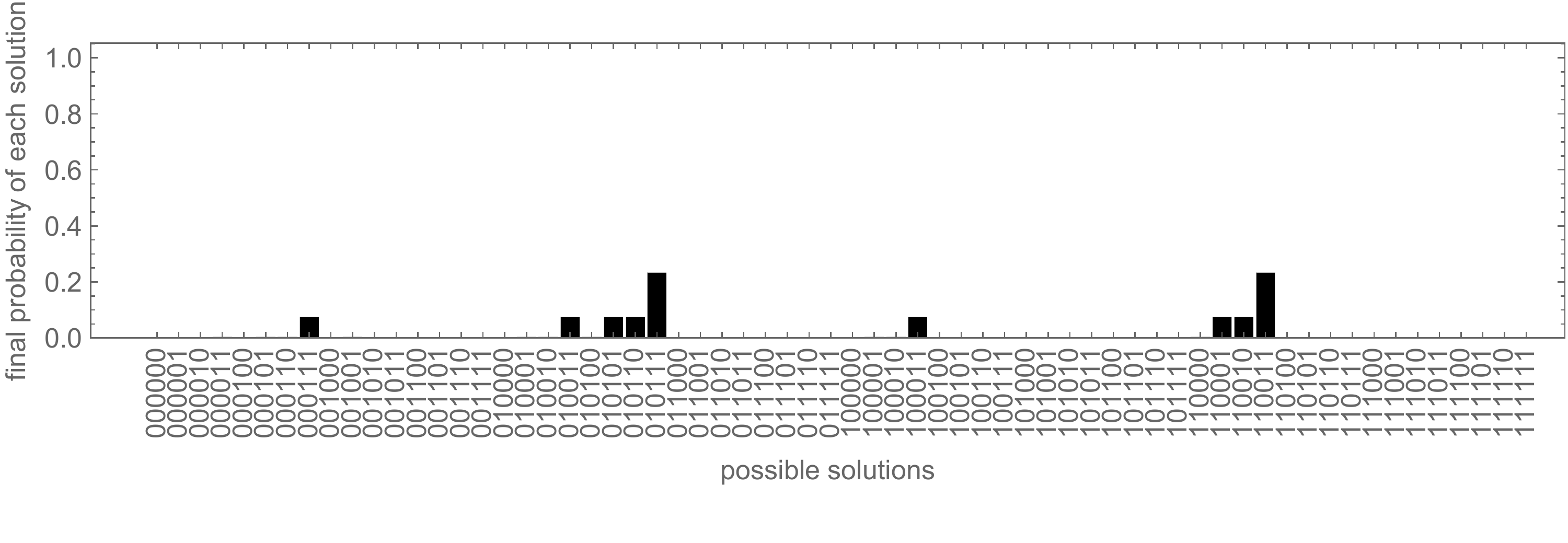}
    }  
    \quad  
    \subfigure[Star graph approach, $\vec{x^*}=(000000)^T$, $p=4$]{  
	\label{Fig-9c} 
    \includegraphics[width=12cm]{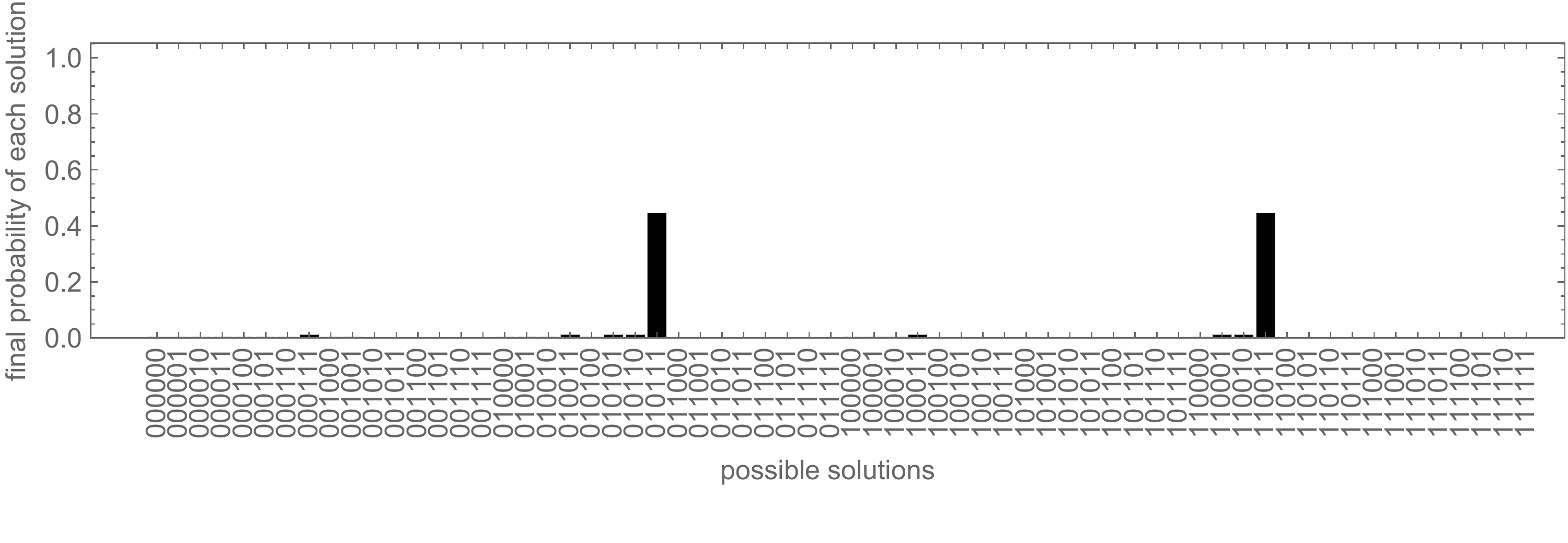}
    }  
    \quad  
     \subfigure[Star graph approach, $\vec{x^*}=(000000)^T$, $p=5$]{ 
	\label{Fig-9d}  
    \includegraphics[width=12cm]{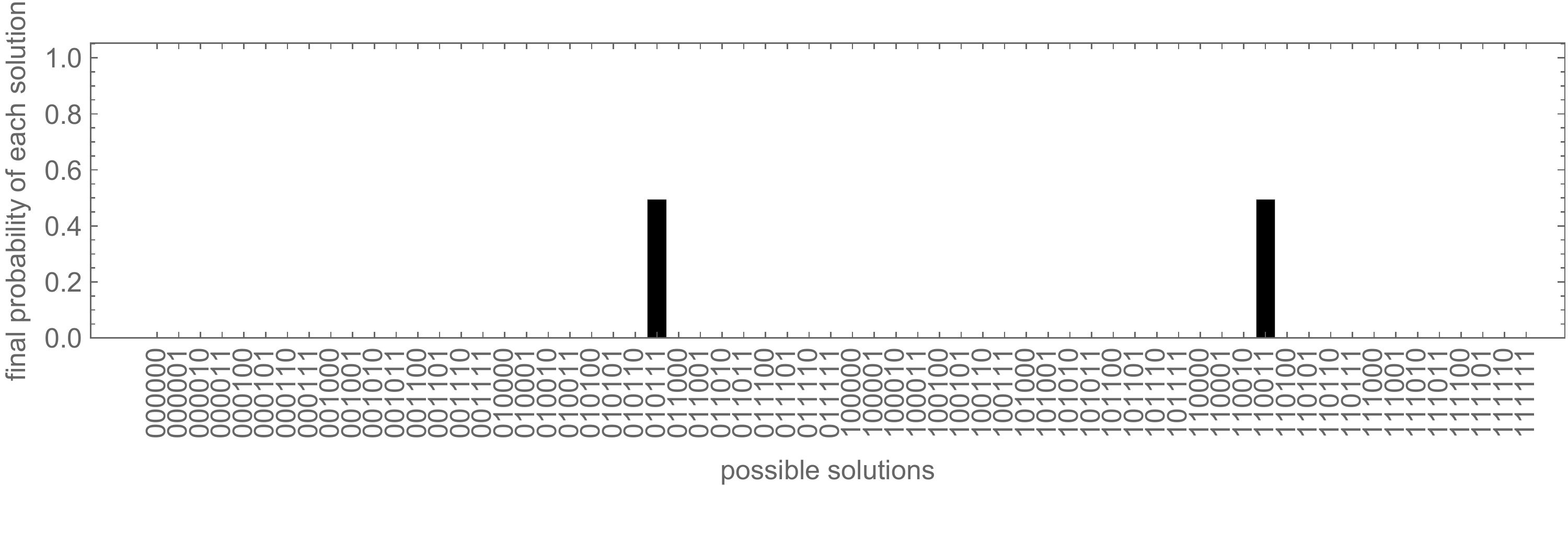}  
    }
    \caption{Results for the Set Packing problem.}
\end{figure}

For this section, we consider Set Packing as the sample problem. Here we present a concrete instance as $S=\{a_1, a_2, a_3, a_4, a_5, a_6, a_7, a_8\}$ and $S_1=\{a_1, a_3\}$, $S_2=\{a_2\}$, $S_3=\{a_4, a_5\}$, $S_4=\{a_2, a_5, a_6\}$, $S_5=\{a_5, a_8\}$, $S_6=\{a_6, a_7\}$. The trivial feasible solution meeting the constraint is ``000000'' and we use this solution as the central vertex in the star graph method. The optimal solutions are ($S_1, S_2, S_3, S_5$) and ($S_1, S_2, S_5, S_6$), which do not have elements in common. The solutions are shown in Fig 9. \cref{eq:bsingleinequality} provides a nearly regular operator which leads to a perfect probability for getting the optimal solution when $p=3$ as shown in \cref{Fig-9a}, while at the same $p$, the irregular star graph operator is considerably less likely to get the optimal solution (\cref{Fig-9b}). But as long as increasing $p$, star graph operator can get the optimal solution with higher probability either (\cref{Fig-9c} and \cref{Fig-9d}).

\subsection{Discussion}
The regularity of constraint encoded $\hat{B}$ could be defined as the difference between the maximum degree and the minimum degree of vertices in the consequential connected subgraph as per \cref{eq:regularity}. The smaller the difference, the more regularity of $\hat{B}$ is. Since all the vertices in our $\hat{B}$ as per \cref{e5} have the same degree, the value of such difference is 0, which denotes the most regularity, therefore produces the highest-quality result. While this value of $\hat{B}$ as per \cref{e6} $\geq 2$ (The value becomes larger when the number of vertices in $\hat{B}$ increases), indicates less regularity than \cref{e5} for getting the optimal solution with less probability under fixed $p$ iteration. Moreover, another operator $\hat{B}$ as per \cref{eq:bsingleinequality} is a tailored transverse field operator whose regularity value $=1$. We name it as the near regular operator, which provides higher quality solutions than others in solving optimization problems with linear inequality constraint. The star graph operator, whose regularity value is $n-2$, has the worst regularity. We use it in the case without alternatives.

\begin{equation}
\label{eq:regularity}
the\ regularity\ of\ \hat{B} = max\_degree(\{\vec{x}\}) - min\_degree(\{\vec{x}\}), \quad where\ \vec{x}\in \Omega
\end{equation}
 
Besides the specific operators we used in the above two subsections, we also change the degree of \cref{e5} manually, such as from 4-regular to 3-regular/5-regular, to see the influence of such change on getting the optimal solution. We observe that influence does not exist at all since we get the precise same final probability distribution. Moreover, substituting other classical optimization methods provided by $Mathematica^{@}$ for NelderMead simplex algorithm, which we use to get the figures of all problem instances in this paper, the QAOA end up with almost identical results with negligible differences in the final probability. All these pieces of evidence manifest \cref{eq:regularity} is a good indicator of the regularity and which is seemingly a positive factor for achieving the higher probability of the optimal solution.

One may wonder, does the advantage obtained from the regularity of the operator $\hat{B}$ is a general feature to all NP optimization problems in the context of QAOA? As we know, the corresponding decision version of almost all NP optimization problems are NP-complete problems, which can be mutually reducible in polynomial time. Hence other NP optimization problems can transform into the verified instances then get such feature. 

So why the regularity of the operator $\hat{B}$ is in favor of achieving a high-quality solution? We get an inspired interpretation from Zhou et al. 's work\cite{zhou2018quantum} and Rezakhani et al. 's work\cite{rezakhani2009quantum}.

Zhou et al. analyzed the relationship between QAOA and QAA, converted QAOA parameters to a well-defined annealing path by the optimal evolution time defined by the \{$\gamma_i$\} and \{$\beta_i$\}\cite{zhou2018quantum}. Rezakhani et al. recast the optimal annealing path (Quantum Adiabatic Brachistochrone, QAB) in a natural differential-geometric framework, indicated the optimal path is a geodesic in the parameter manifold embedded in the Riemann space\cite{rezakhani2009quantum}.

QAOA can be reconsidered from the perspective of differential geometry. The optimized parameters of QAOA, namely \{$\gamma_i$\} and \{$\beta_i$\}, constitute a low dimensional manifold to approximate the original parameter space (in analogy of \cite{rezakhani2009quantum}). The lower the dimension of this manifold (smaller $p$, a smaller number of $\gamma_i$ and $\beta_i$), the lower degree of approximating the original parameter space, thus the smaller probability of getting the optimal solution. This perspective is consistent with our numerical experiment, and also consistent with the original theory, as we know only if $p\rightarrow \infty$, the QAOA can approximate QAA perfectly.

However, infinite-dimensional parameter space (infinite $p$) does not make sense. We need to use limited $p$, i.e. a limited number of optimized $\gamma_i$ and $\beta_i$ to construct a geodesic in the low dimensional manifold to achieve the nearly optimal solution. The observation in our numerical experiments indicates that the regularity of the operator $\hat{B}$ is in favor of getting a lower-dimensional manifold to approximate the original space, i.e., the conjecture --- In the context of using QAOA to solve combinational optimization problems with constraints, if the corresponding graph of the constraint encoded $\hat{B}$ has more regularity, then it is in favour of achieving a higher probability of the optimal solution than other constraint encoded $\hat{B}$ with less regularity under the condition of fixed p iteration.

Although we deem that the regular $\hat{B}$ has good property in the context of QAOA, it is hard to quantify the advantage for comparison, because the computation of curvature tensor of an arbitrary Hermitian operator (theoretically, any Hermitian operator could be constraint encoded $\hat{B}$) is too complicated. But we know that in fact, the symmetric property of an entity is a positive factor to simplify its representation, computation, etc. For example, the quadratic curve on a plane requires two-dimensional parameters $x$ and $y$ to represent in general. However, the most symmetric quadratic curve, the circle can be represented as a manifold in one dimension, which only needs one parameter $radius$ to represent in polar coordinates. In QAOA $\hat{C}$ is a diagonal matrix, the ``shape'' of $\hat{B}$ should be the crucial factor in the computation of the instantaneous Hamiltonian, which determines the optimal adiabatic path. If $\hat{B}$ is symmetric (regular), then the instantaneous Hamiltonian should be more symmetric and in other words, can be probably simplified.

\section{Encoding constraints in $\hat{C}$}

Finally, we discuss the alternative approach that keeps the $\hat{B}$ operator as the unmodified transverse field Hamiltonian, and instead modifies the $\hat{C}$ operator to encode the constraint. In general, $\hat{C}$ takes the form $\sum_x w_x|x\rangle\langle x|$, where $w_x$ is the solution cost or quality corresponding to the combinatorial solution $\vec{x}$. This operator is diagonal under the computational basis. As shown in \cref{Sec:Prerequisite}, we can get $\Omega$ containing all feasible solutions (\cref{eq:omega}). Hence the projection operator $M=\sum_{x \in \Omega}|x\rangle\langle x|$ can also be obtained. If we then perform the operation $M^\dagger \hat{C} M$, the constraint-encoded operator $\hat{C}$ is produced.

Using the constraint-encoded operator $\hat{C}$ is plausible, and does not require penalty terms. But, since the operator works in the entire Hilbert space, the final result is generally not as good as by constraining the subspace with $\hat{B}$. Fig. 10 shows the result of applying this method to QAOA with $p=3$, in order to find a solution to the graph partition problem on \cref{Fig1-instance} (studied in \cref{Sec:GP} \cref{Fig: probability1} and \cref{Fig: probability2}). There is a significant probability of measuring low-quality solutions compared to the $\hat{B}$-modification schemes. The result of this experiment agrees with Hadfield's work \cite{hadfield2017qantum}, where it is argued that decreasing the size of the considered subspace leads to improved QAOA results.

\begin{figure}[!htbp]  \label{fig:C}
    \includegraphics[width=11cm]{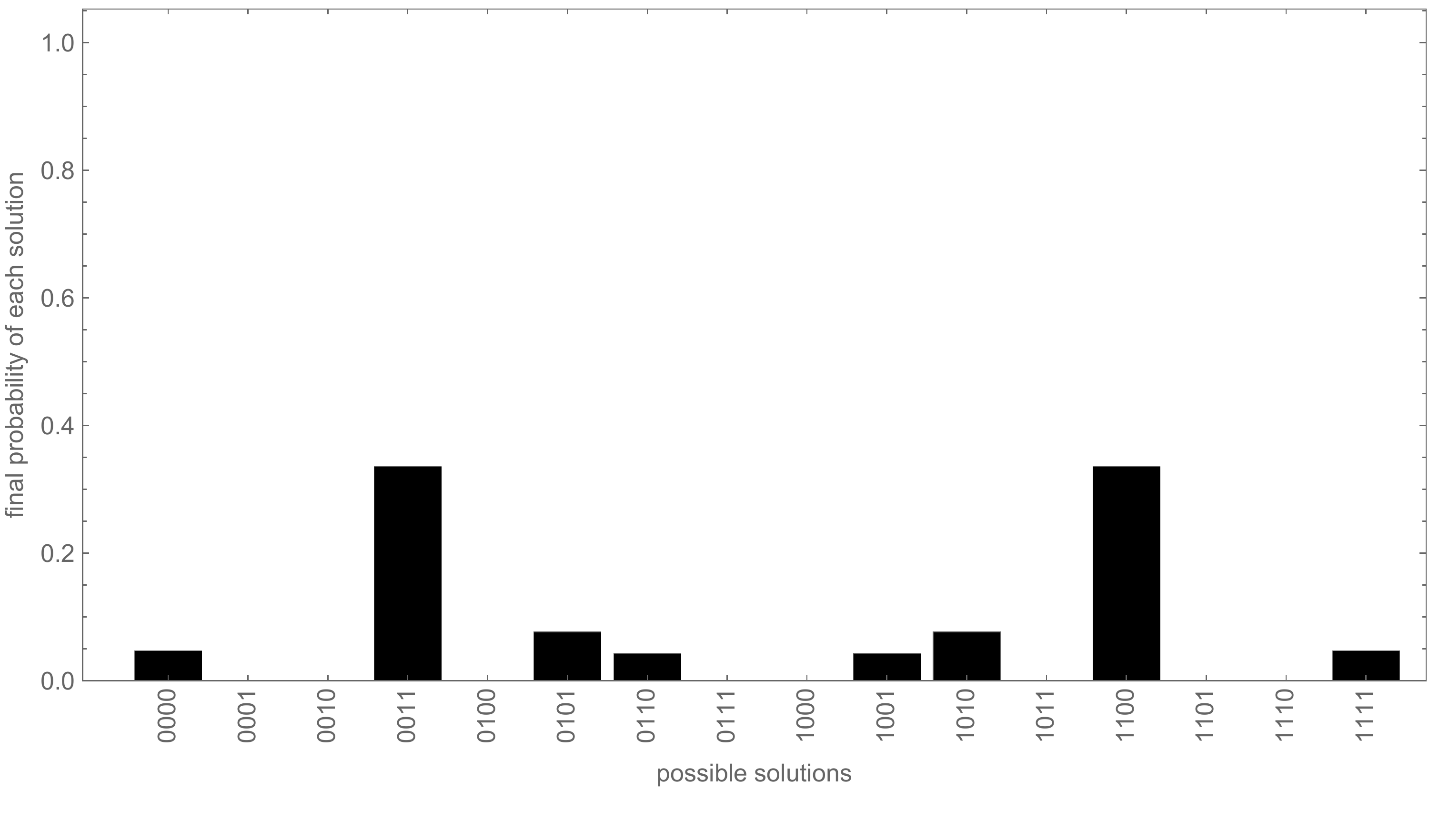}  
    \caption{Results for the graph partition problem in Fig. 1, solved by $p=3$ QAOA.}
\end{figure}

\section{\label{sec:level1}Conclusion}

In this paper, we established a generalized scheme to deal with constraints when solving constrained combinational optimization problems with QAOA. We classified the constraints into three categories -- linear equality constraint, linear inequality constraint, and arbitrary constraints. The linear constraint can be encoded uniformly in operator $\hat{B}$ using 
\begin{equation}
\hat{B}=\sum\limits_{\substack{\vec{x}, \vec{x'} \in \Omega \\ d(\vec{x},\vec{x'})=1\ or\ 2}}\ket{x}\bra{x'}+\ket{x'}\bra{x}
\end{equation}
where $d(\vec{x},\vec{x'})$ is the Hamming distance between two feasible solutions $\vec{x}$ and $\vec{x'}$. When dealing with arbitrary constraints, we used a star graph as the encoding approach, where an arbitrary feasible solution is chosen to be the central vertex. We applied this scheme to the optimization problems to a variety of NP optimization problems. The results and comparison with other schemes demonstrated its effectiveness and efficiency for solving constrained optimization problems with QAOA, albeit for the small problem instances that can be analyzed with classical computers.

We consider QAOA as a high-performing algorithm for resolving optimization problems. However, QAOA combined with the hybrid quantum-classical variational optimization scheme can only be considered a heuristic. Theoretically, QAOA is the approximation of QAA only produces the optimal result with certainty when $p\rightarrow\infty$ \cite{farhi2014quantum}. However, this paper reinforces the observation made in \cite{farhi2014quantum}, that a small value $p$ appears adequate to obtain the optimal (or at least, a high-quality) solution. If this is indeed the case for large problem instances, QAOA is a powerful utility that can be applied to a wide range of real-world optimization problems to efficiently obtain high-quality approximate solutions.

\section{Acknowledgements}
This work is supported by the Natural Science Foundation, China (Grant No.61802002) and Natural Science Foundation of Anhui Province, China (Grant No.1708085MF162) .

\bibliographystyle{unsrt}  
\bibliography{ref} 

\begin{thebibliography}{10}
\bibitem{difference}
{AQC} is to used for searching minimum. {QAOA} is used for searching maximum.
  {T}he operator $\hat{C}$ defined here differs slightly from the original
  definition proposed by {I}. {H}en.
\end{thebibliography}

\end{document}